\documentclass[letterpaper]{article} 
\usepackage{aaai2027} \nocopyright 
\usepackage[hyphens]{url} 
\usepackage{graphicx} 
\usepackage{natbib} 
\usepackage{caption} 
\usepackage{amsmath}
\usepackage{amssymb}
\usepackage{booktabs}
\usepackage{multirow}
\usepackage{algorithm}
\usepackage{algpseudocode}

\title{Benign on Label, Malicious by Design: Clean-Label Dormant-to-Activated Backdoor via Machine Unlearning with Removable Camouflage}

\author{
\normalfont
Dongdong Zhao\textsuperscript{\rm 1},
Can Li\textsuperscript{\rm 1},
Xiang Yao\textsuperscript{\rm 1},
Fan He\textsuperscript{\rm 1},
Qihang Ge\textsuperscript{\rm 1},
and Baogang Song\textsuperscript{\rm 1,*}
}

\affiliations{
\textsuperscript{\rm 1}\textit{Wuhan University of Technology}\\
{\small\textit{Email: zdd@whut.edu.cn, 375291@whut.edu.cn, yaoxiang@whut.edu.cn, hefan@whut.edu.cn, 367999@whut.edu.cn, 297710@whut.edu.cn}}
}

\begin{document}

\maketitle
\begin{abstract}

Existing backdoor attacks often become effective immediately after backdoor implantation and may therefore be exposed before exploitation. Machine unlearning activated dormant backdoors mitigate such behavioral exposure by remaining inactive after training and becoming effective only after selected training records are unlearned. However, existing methods struggle to simultaneously achieve a low pre-unlearning attack success rate and strong post-unlearning activation under clean-label constraints and realistic unlearning requests.
Achieving this transition requires jointly establishing a persistent latent association and a removable suppressive influence.
To address this challenge, we propose a clean-label unlearning-activated backdoor framework based on dual-generator learning and formulate it as a bilevel optimization problem: By simulating latent backdoor establishment and machine unlearning, the framework alternately learns sample-specific triggers that establish a latent trigger-to-target association and label-consistent camouflage samples that provide removable suppression. Once a small subset of camouflage samples is unlearned, the suppression is lifted and the dormant backdoor is activated. Experiments on CIFAR-10 and ImageNet-10 show that our method maintains lower pre-unlearning attack success rates while achieving stronger post-unlearning activation across multiple unlearning algorithms than representative backdoor baselines.
These results demonstrate that reliable dormancy-to-activation transitions can be achieved by coordinating a persistent latent association with removable suppression under clean-label and realistic deletion constraints.

\end{abstract}

\section{Introduction}
\begin{figure}[!t]
    \centering
    \includegraphics[
        width=0.98\columnwidth
    ]{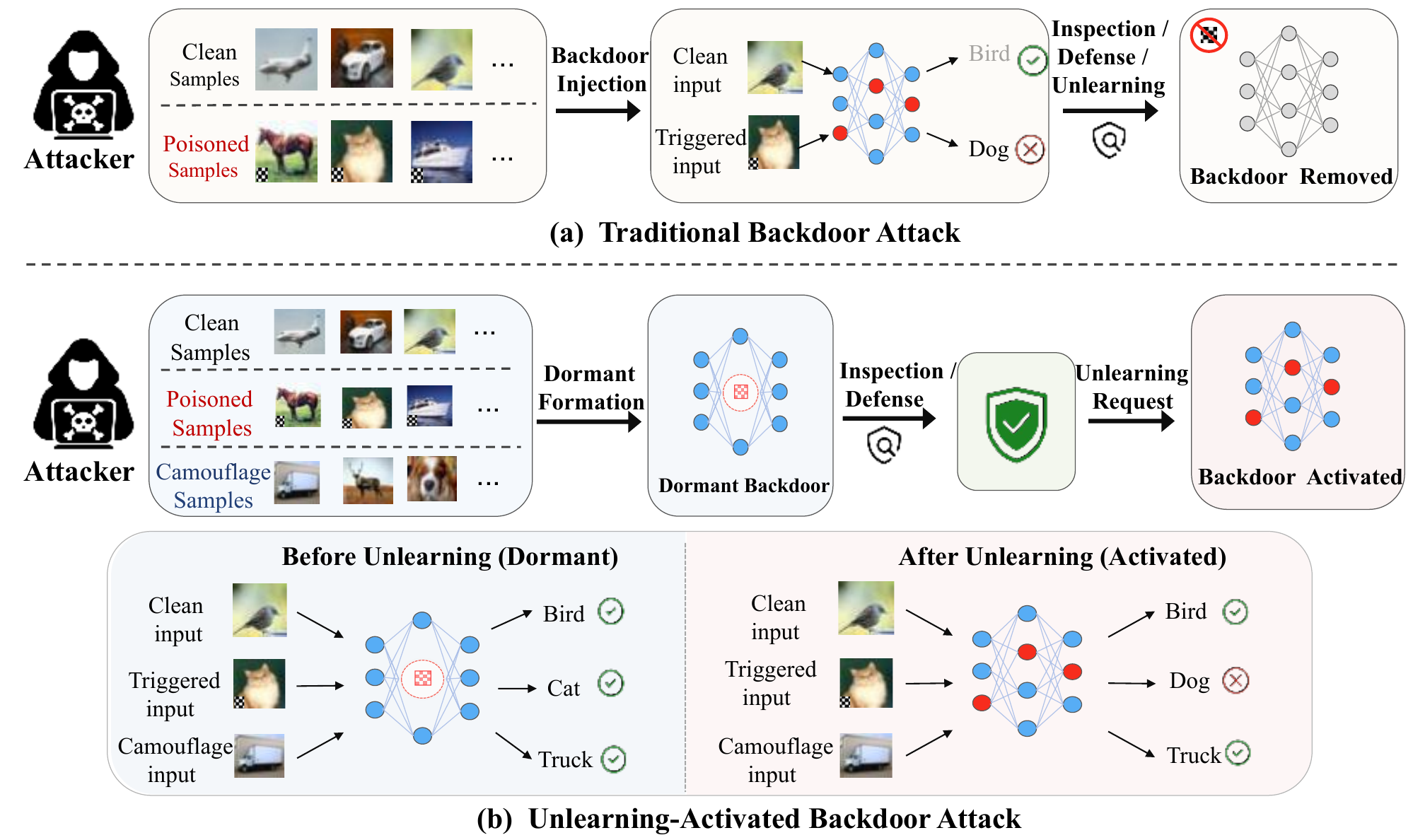}
    \caption{Research motivation.}
    \label{fig:motivation}
\end{figure}

Backdoor attacks threaten DNNs deployed in safety-sensitive applications by preserving benign performance while inducing attacker-specified predictions on triggered inputs. 
 Existing studies improve stealthiness through invisible, warping-based, input-adaptive, and clean-label designs \cite{li2021invisible,nguyen2021wanet,nguyen2020inputaware,nguyen2025wicked}, mainly reducing visual or label-level anomalies in poisoned samples. 
However, these improvements usually do not change a critical fact: the learned backdoor often becomes active immediately after training.
Thus, malicious trigger-to-target behavior may be exposed by post-training inspection or backdoor defenses before the attacker exploits it, as shown in Fig.~\ref{fig:motivation}~\cite{wang2019neuralcleanse,gao2019strip}.
This reveals a behavioral exposure risk that cannot be fully addressed by merely making poisoned samples less perceptible.

To reduce such post-training exposure, recent studies have explored dormant or delayed-activation backdoors that remain inactive after training and are activated only by a specific subsequent event. 
Machine unlearning provides a natural post-training activation mechanism by removing the influence of specified training samples in response to deletion requests, thereby inducing model-state changes that attackers can exploit to activate a dormant backdoor.
Existing studies have demonstrated the feasibility of unlearning-activated backdoors, but important limitations remain in realistic data-submission and deletion-request scenarios.
Some rely on dirty-label poisoning, violating label consistency
\cite{zhang2023exploiting,alam2025reveil}.
Some construct influence-driven suppression for a given off-the-shelf backdoor construction,
rather than jointly constructing a latent association and its suppression mechanism under clean-label constraints
\cite{huang2024uba}. 
Moreover, the limited work under clean-label constraints activates the backdoor by unlearning samples outside the attacker-submitted training records
\cite{unclean}.
This deviates from practical machine-unlearning semantics, where a valid request should be record-matched and restricted to data submitted by the requester for target-model training.
Therefore, although existing unlearning-activated backdoors mitigate the immediate post-training exposure of backdoor behavior, they do not fully satisfy the dual considerations of clean-label data submission and realistic user deletion requests.

Motivated by these limitations, we study a stricter clean-label data-removal workflow in which all attacker-crafted samples retain their original labels and the forget set is restricted to attacker-submitted samples that actually participated in training. This workflow jointly enforces label consistency at data submission and sample provenance at deletion, better reflecting realistic machine-unlearning services. Under this workflow, the core challenge is to simultaneously achieve label consistency, pre-unlearning dormancy, and post-unlearning activation. Specifically, the trigger-target association must be implanted implicitly under original-label supervision, remain unexposed after training, and emerge only after a subset of attacker-submitted samples is forgotten. To satisfy these requirements, the attack framework needs to coordinate two types of training signals: one that establishes a latent trigger-target association that persists through unlearning, and another that suppresses this association before unlearning through attacker-submitted samples that can later be deleted. Coordinating these two effects under clean-label constraints is the key challenge for realizing clean-label unlearning-activated backdoors across pre- and post-unlearning model states.

To address this challenge, we propose a dual-generator framework that separately models latent backdoor implantation and camouflage suppression and coordinates them through bilevel optimization. This separation is necessary because establishing a persistent trigger-target association and suppressing its observable behavior impose competing objectives that may interfere when optimized through a single generator. Accordingly, the trigger generator produces sample-specific triggers that establish a latent association under original-label supervision and preserve it through unlearning, while the camouflage generator produces label-consistent samples that impose removable suppression on the trigger-induced target tendency during training. Since unlearning-activated backdoors depend on both the latent state after training and the activated state after machine unlearning, we use a bilevel formulation to characterize the coupling between generator optimization and the training--unlearning process. After generator optimization, both trigger-bearing samples and camouflage samples are incorporated into target-model training with their original labels. When the attacker subsequently requests unlearning of a small subset of camouflage samples, the removable suppression is lifted, allowing the retained latent trigger-target association to emerge as an activated backdoor.

Our main contributions are summarized as follows:
\begin{itemize}
\item We systematically study unlearning-activated backdoor attacks under a clean-label data-removal workflow, extending dormant backdoors to a more practical scenario that simultaneously enforces label consistency and realistic user-deletion constraints.

\item We design a bilevel optimization-based dual-generator framework that jointly optimizes latent trigger persistence and removable camouflage suppression by simulating dormant backdoor formation and unlearning-induced activation.

\item We design removable camouflage objectives that combine gradient opposition and target-tendency suppression, enabling label-consistent suppression of trigger-induced behavior that can be lifted through unlearning.

\item We evaluate the effectiveness of our method on CIFAR-10 and ImageNet-10.
Results show that our method effectively realizes clean-label unlearning-activated backdoor attacks, achieving stealthier pre-unlearning dormancy and more effective post-unlearning activation than mainstream backdoor attacks.
\end{itemize}

\section{Related Work}

\subsection{Machine Unlearning}
Machine unlearning aims to remove the influence of requested training samples from a trained model, motivated by privacy regulations and data deletion demands~\cite{otto2018regulation}.
Prior work has developed a range of unlearning methods, such as SISA~\cite{bourtoule2021machine}, First-Order unlearning~\cite{warnecke2023machine}, and PUMA unlearning~\cite{wu2022puma}.
These studies mainly focus on improving unlearning efficiency or preserving model utility. In contrast, we examine unlearning from an adversarial perspective, exploiting the induced model transition as an attacker-controlled mechanism to remove camouflage suppression and activate a dormant clean-label backdoor.
\subsection{Backdoor Attacks}

Backdoor attacks poison model training to implant trigger-specific malicious behavior while preserving benign performance~\cite{gu2019badnets}. Although subsequent invisible, adaptive, warping-based, and clean-label designs improve visual or label-level stealthiness, their backdoors are typically active immediately after training and may be exposed before exploitation. Dormant and delayed-activation attacks instead postpone malicious behavior until a subsequent model-state change, with prior studies showing that deletion-induced updates can activate hidden backdoors~\cite{zhang2023exploiting,liu2024backdoor}. However, these approaches generally rely on dirty-label poisoning or label manipulation.
More recent methods improve activation controllability under machine unlearning. UBA-Inf strengthens the pre-unlearning suppression of an off-the-shelf backdoor using influence functions~\cite{huang2024uba}, but does not jointly construct the latent association and its suppression under clean-label constraints. UNCLEAN studies a clean unlearning attack by amplifying a weak malicious signal through forgetting the clean counterparts of poisoned samples~\cite{unclean}, while its forget set is not restricted to attacker-submitted training records. In contrast, we jointly optimize label-consistent trigger samples and removable camouflage samples under a clean-label data-removal workflow, enabling the camouflage samples to suppress the latent association before unlearning and release it after a valid attacker-controlled deletion request.
\section{Problem Statement}
\noindent\textbf{Problem Formulation.}
We formally define the \emph{Clean-Label Unlearning-Activated Backdoor Attack} as follows.
Let $f_\theta$ denote a classification model trained on 
$\mathcal{D}=\{(x_i,y_i)\}_{i=1}^{N}$, with $x_i\in\mathcal{X}$ and $y_i\in\mathcal{Y}$. 
Here, $\mathcal{X}$ and $\mathcal{Y}$ denote the input and label spaces, respectively.
The attacker aims to implant a clean-label latent backdoor that remains inactive after normal training and becomes activated once camouflage samples are unlearned.
Specifically, the attacker first selects a target label $y_t\in\mathcal{Y}$ and controls a limited subset $\mathcal{D}_a\subset\mathcal{D}$ of the training data.
From $\mathcal{D}_a$, the attacker selects a backdoor source subset
$\mathcal{P}\subset
\{(x,y)\in\mathcal{D}_a\mid y=y_t\}$ with poisoning rate $\rho_p$, and a camouflage source subset
$\mathcal{C}\subset\{(x,y)\in\mathcal{D}_a\mid y\neq y_t\}$ with rate $\rho_c$.
Based on the trigger generator $G_\phi$ and the camouflage generator $M_\psi$, the attacker constructs the backdoor sample set $\mathcal{P}'=\{(B_\phi(x),y)\mid (x,y)\in\mathcal{P}\}$ and camouflage sample set $\mathcal{C}'=\{(C_{\phi,\psi}(x),y)\mid (x,y)\in\mathcal{C}\}$, respectively, where the detailed formulations of $B_\phi(\cdot)$ and $C_{\phi,\psi}(\cdot)$ are provided in the Method section.
All crafted samples preserve their original labels.
The crafted samples are submitted through the normal data collection pipeline and 
their incorporation results in the
mixed training set
$\mathcal{D}^{*}=(\mathcal{D}\setminus\mathcal{D}_a)
\cup(\mathcal{D}_a\setminus(\mathcal{P}\cup\mathcal{C}))
\cup\mathcal{P}'\cup\mathcal{C}'$,
which is subsequently used by the model owner for training.
After training on $\mathcal{D}^{*}$, the parameters of the model are updated to $\theta_l$, resulting in a latent backdoor model $f_{\theta_l}$ that behaves normally on clean input and remains inactive on triggered inputs before unlearning:
\begin{equation}
f_{\theta_l}(x_i)=y_i,\qquad 
f_{\theta_l}(B_\phi(x_i))=y_i
\end{equation}
At the intended attack time, the attacker requests unlearning of
$\mathcal{U}\subseteq\mathcal{C}'$. 
The model owner then invokes an unlearning algorithm to update the model parameters from $\theta_l$ to $\theta_u$, obtaining the unlearned model $f_{\theta_u}$. 
After unlearning, the attack aims to activate the backdoor and induce the target behavior:
\begin{equation}
f_{\theta_u}(x_i)=y_i,\qquad 
f_{\theta_u}(B_\phi(x_i))=y_t
\end{equation}
\paragraph{Threat Model.}
We assume that the attacker participates in data collection and can contribute a limited number of crafted backdoor and camouflage samples to the model owner's training set.
During offline optimization, the attacker trains surrogate models to approximate the victim training and unlearning processes. 
The attacker can later access the unlearning interface of the data provided by the model service provider and request the removal of a small subset of training samples. 
However, the attacker cannot interfere with victim training,
directly modify model parameters, or control final model
deployment.
All crafted samples retain their original labels, reflecting realistic settings where attackers have limited label-control ability and mislabeled records are more readily detected during data inspection or deletion verification.

\section{Method}
In this section, we first present the overall attack workflow and then
elaborate on the optimization of the two generators. 

\subsection{Overview}
\begin{figure*}[!t]
    \centering
    \includegraphics[
        width=0.92\textwidth
    ]{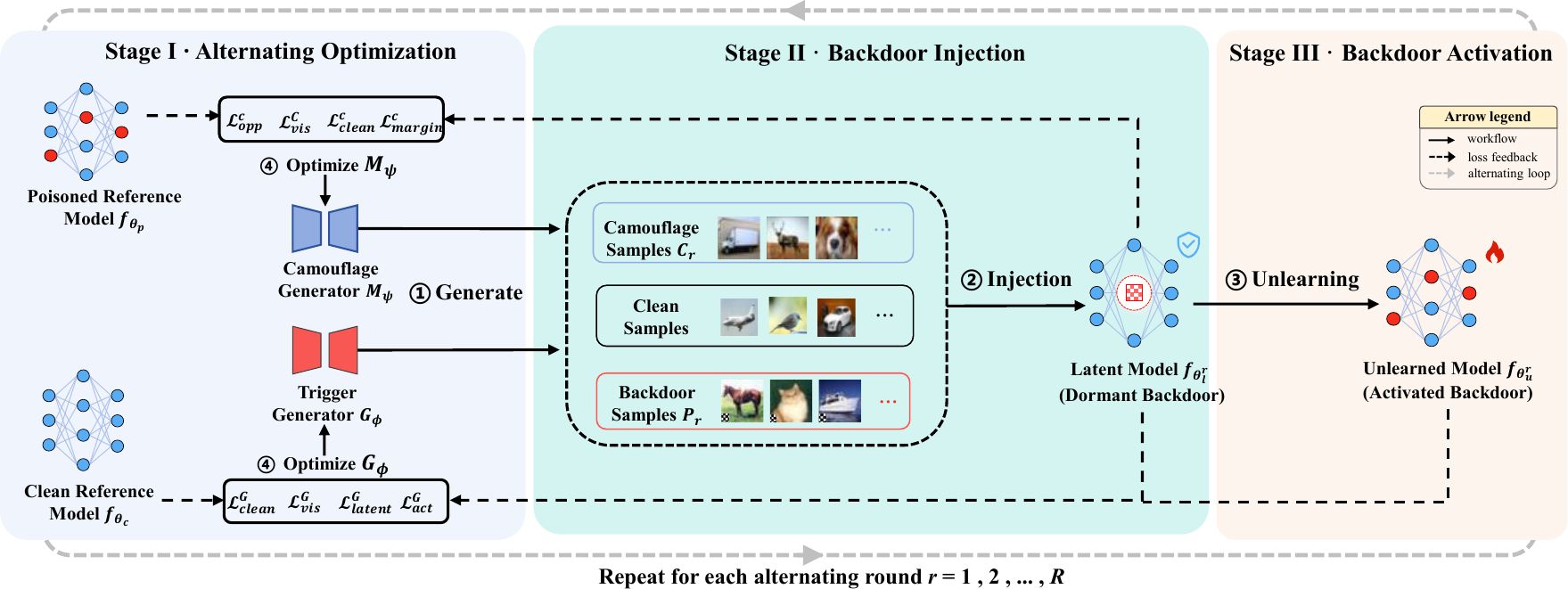}
    \caption{Overview of the proposed framework.}
    \label{fig:framework}
\end{figure*}

As illustrated in Fig.~\ref{fig:framework}, our proposed unlearning-activated backdoor attack proceeds in three stages.
\begin{enumerate}
    \item \textbf{Dual-Generator Optimization.}
The attacker optimizes the trigger generator and the camouflage generator under a bilevel formulation in an offline manner. 
The former learns a latent trigger association, while the latter learns a removable suppressive influence.

    \item \textbf{Latent Backdoor Establishment.}
After offline optimization, the generated backdoor and camouflage samples are injected into the training set, causing the victim model to learn a dormant backdoor whose
malicious behavior remains suppressed after deployment.

    \item \textbf{Backdoor Activation.}
At the intended attack time, the attacker requests the unlearning of the camouflage samples, removing their suppressive influence and activating the latent backdoor.
\end{enumerate}
To realize the above three-stage workflow and the controlled
dormant-to-activated transition, we next introduce the
construction and optimization objectives of $G_\phi$ and
$M_\psi$, followed by their bilevel optimization.

\subsection{Trigger Generation and Optimization}
\paragraph{Backdoor Sample Construction.}
We formulate backdoor sample construction as a learnable
injection process, where $G_\phi$ produces a sample-specific
perturbation $G_\phi(x)$ for each input $x$.
To reduce high-frequency artifacts and spatial
discontinuities, we project the generated perturbation onto
the low-frequency domain.
We define the low-frequency projection operator as~\cite{ahmed1974discrete}
\begin{equation}
\mathcal{P}_{lf}(z)
=
\mathrm{IDCT}
\left(
\mathcal{M}_{lf}\odot \mathrm{DCT}(z)
\right)
\label{eq:lf_projection}
\end{equation}
where $\mathcal{M}_{lf}$ denotes the low-frequency mask and $\odot$ denotes the Hadamard product.
The filtered trigger perturbation is then obtained as $\delta_G(x)=\mathcal{P}_{lf}(G_\phi(x))$.
To ensure valid pixel values and further improve the spatial continuity and visual stealthiness of the trigger, we inject the low-frequency perturbation into the original image, followed by clipping and Gaussian smoothing~\cite{huynh2024combat}:
\begin{equation}
B_{\phi}(x)
=
\mathcal{K}
\left(
\mathrm{Clip}
\left(
x+\eta_G\cdot \delta_G(x)
\right)
\right)
\label{eq:backdoor_sample}
\end{equation}
where $\eta_G$ controls the trigger strength,
$\mathrm{Clip}(\cdot)$ ensures valid pixel values, and
$\mathcal K(\cdot)$ denotes Gaussian smoothing.

\paragraph{Trigger Generator Optimization.}
In the unlearning-activated backdoor attacks, the core challenge of trigger optimization lies in the fact that the trigger should exhibit different behaviors before and after unlearning. 
Unlike conventional backdoor attacks, which only require the trigger to be effective after model training, our trigger should remain dormant in the normally deployed model and become effective after camouflage samples are unlearned. 
Therefore, when optimizing the trigger generator, we explicitly consider both the latent model $f_{\theta_l}$ and the unlearned model $f_{\theta_u}$.
In the trigger generator optimization, we consider the following loss terms:
\begin{itemize}
    \item \textbf{Latent loss.}
    Before unlearning, the latent loss requires triggered samples to be classified into their ground-truth labels:
    \begin{equation}
    \mathcal{L}_{latent}^{G}
    =
    \mathbb{E}_{(x,y)\in\mathcal{D}}
    \mathcal{L}(f_{\theta_l}(B_\phi(x)),y)
   \end{equation}
    This loss constrains the generated triggers to remain dormant in the latent model by preventing premature target behavior before unlearning.

    \item \textbf{Activation loss.}
    For activation, we apply the trigger to samples in the dataset and require the unlearned model to classify the triggered samples into the target label $y_t$:
    \begin{equation}
    \mathcal{L}_{act}^{G}
    =
    \mathbb{E}_{(x,y)\in\mathcal{D}}
    \mathcal{L}(f_{\theta_u}(B_\phi(x)),y_t)
    \end{equation}
    This objective makes the trigger effective after unlearning, enabling the attacker to achieve the intended attack.
\end{itemize}

\begin{itemize}

    \item \textbf{Visibility loss.}
    We constrain the perturbation magnitude to improve the visual stealthiness of the trigger:
    \begin{equation}
    \mathcal{L}_{vis}^{G}
    =
    \mathbb{E}_{x\in\mathcal{D}}
    \|B_\phi(x)-x\|_2^2
    \end{equation}

    \item \textbf{Non-adversarial loss.}
    We require the clean model $f_{\theta_c}$ to correctly classify the generated backdoor samples, preventing the trigger from degenerating into a standard adversarial perturbation:
    \begin{equation}
    \mathcal{L}_{clean}^{G}
    =
    \mathbb{E}_{(x,y)\in\mathcal{D}}
    \mathcal{L}(f_{\theta_c}(B_\phi(x)),y)
    \end{equation}

\end{itemize}
Combining the above objectives, the trigger generator is trained to solve the following optimization problem:
\begin{flalign}
G_\phi^{*}
&=
\arg\min_{G_\phi}
\left\{
\begin{aligned}
&\lambda_l\mathcal{L}_{latent}^{G}
+\lambda_a\mathcal{L}_{act}^{G} \\
&+\lambda_v\mathcal{L}_{vis}^{G}
+\lambda_c\mathcal{L}_{clean}^{G}
\end{aligned}
\right. && \label{eq:trigger_opt} \\
\text{s.t.}\quad
\theta_l
&=
\arg\min_{\theta}
\mathbb{E}_{\mathcal{D}^{*}}
\mathcal{L}(f_{\theta}(x),y), && \notag \\
\theta_u
&=
\mathcal{F}_{\mathrm{unlearn}}(\theta_l,\mathcal{U}) && \notag
\end{flalign}
where $\lambda_l$, $\lambda_a$, $\lambda_v$, and $\lambda_c$ are balancing coefficients, 
$\mathcal{F}_{\mathrm{unlearn}}$ denotes the machine unlearning algorithm, and 
$\mathcal{U}$ is the requested forget set.
\subsection{Camouflage Generation and Optimization}
\paragraph{Camouflage Sample Construction.}
Given an input image $x$ from the camouflage set, the current
trigger generator $G_\phi$ first produces a raw trigger
perturbation, which is projected onto the low-frequency
components using Eq.~\eqref{eq:lf_projection}:
$\delta_G(x)=\mathcal{P}_{lf}(G_\phi(x))$.
We then construct a trigger-conditioned intermediate sample
as
\begin{equation}
x^d
=
\mathrm{Clip}
\left(
x+\eta_G\cdot\delta_G(x)
\right)
\label{eq:intermediate_sample}
\end{equation}
Conditioned on $x^d$, the camouflage generator $M_\psi$
produces a camouflage perturbation, whose low-frequency
form is
$\delta_C(x^d)=\mathcal{P}_{lf}(M_\psi(x^d))$.
The final camouflage sample is constructed as
\begin{equation}
C_{\phi,\psi}(x)
=
\mathcal{K}
\left(
\mathrm{Clip}
\left(
x^d+\eta_C\cdot\delta_C(x^d)
\right)
\right)
\label{eq:camouflage_sample}
\end{equation}
where $\eta_C$ controls the camouflage perturbation strength.
\paragraph{Camouflage Generator Optimization.}
In this framework, the key challenge of camouflage generator
optimization is to make camouflage samples exert a removable
suppressive influence. During training, camouflage samples
retain their original labels while suppressing backdoor behavior,
thereby keeping the implanted backdoor inactive before
unlearning. Once these samples are unlearned, their suppressive
influence is removed, allowing the latent association to emerge
as an activated backdoor. We optimize the camouflage generator
using the following loss terms:
\begin{itemize}

\item \textbf{Gradient-opposing loss.}
This loss requires the training gradient of camouflage samples to be anti-aligned with a trigger-induced target direction.
For each camouflage source sample $x\in\mathcal{C}$, we first construct its trigger-contaminated intermediate sample $x^d$.
Rather than deriving this direction from the backdoor sample set $\mathcal{P}'$, which may introduce class-semantic and visual-content mismatch with the current camouflage sample, we use $x^d$ to characterize the trigger-induced target tendency under the same source image:
\begin{equation}
g_{bd}
=
\nabla_{\theta_l}
\mathcal{L}(f_{\theta_l}(x^d),y_t)
\end{equation}
    For the corresponding camouflage sample $C_{\phi,\psi}(x)$, we compute its training gradient:
    \begin{equation}
    g_c
    =
    \nabla_{\theta_l}
    \mathcal{L}(f_{\theta_l}(C_{\phi,\psi}(x)),y)
    \end{equation}
    We define their cosine similarity as
    \begin{equation}
    \operatorname{cos}_{sim}(g_c,g_{bd})
    =
    \frac{
    \langle g_c,g_{bd}\rangle
    }{
    \|g_c\|_2\|g_{bd}\|_2
    }
    \end{equation}
    The gradient-opposing loss is then formulated as
    \begin{equation}
    \mathcal{L}_{opp}^{C}
    =
    1+\operatorname{cos}_{sim}(g_c,g_{bd})
    \end{equation}
   Minimizing $\mathcal{L}_{opp}^{C}$ drives $\operatorname{cos}_{sim}(g_c,g_{bd})$ toward $-1$, causing camouflage gradients to oppose the trigger-induced target direction defined by $x^d$.
This creates a removable suppressive influence before unlearning; once the camouflage samples are unlearned, the gradient cancellation is removed and latent association is released.

   \item \textbf{Target-suppressive margin loss.}
    Since camouflage samples are generated from
trigger-conditioned inputs, they may retain residual
target-class tendency under a model that has learned the
trigger-target association. 
To provide a stable reference for suppressing this residual tendency, we offline construct a trigger-only reference model \(f_{\theta_p}\) before optimization. 
Specifically, we first train an initial trigger generator and use the resulting trigger-bearing clean-label samples, without introducing camouflage samples, to train \(f_{\theta_p}\). 
 We then fix $f_{\theta_p}$ and use it
only for the target-suppressive margin constraint.
    Let $z_k^p(\cdot)$ denote the pre-softmax logit of class $k$ predicted by $f_{\theta_p}$ and define
    \begin{equation}
    s_t(x)=z_{y_t}^{p}(C_{\phi,\psi}(x)), 
    \quad
    s_y(x)=z_y^{p}(C_{\phi,\psi}(x))
    \end{equation}
    The margin loss is defined as
    \begin{equation}
\mathcal{L}_{margin}^{C}
=
\mathbb{E}_{(x,y)\in\mathcal{C}}
\max(0,s_t(x)-s_y(x)+\gamma)
\label{eq:margin_loss}
\end{equation}
    
    where $\gamma$ is the margin hyperparameter.
    Minimizing this loss enforces the original-class logit to exceed the target-class logit by at least $\gamma$, thereby suppressing the target-class tendency of camouflage samples.
\end{itemize}

\begin{itemize}
\item \textbf{Label-preserving loss.}
We require the latent model to classify camouflage samples
as their original labels, preventing suppression through
misclassification:
\begin{equation}
\mathcal{L}_{clean}^{C}
=
\mathbb{E}_{(x,y)\in\mathcal{C}}
\mathcal{L}
\left(
f_{\theta_l}(C_{\phi,\psi}(x)),y
\right)
\end{equation}

    \item \textbf{Visibility loss.}
We constrain camouflage perturbations with an $\ell_1$ penalty to improve visual stealthiness:
\begin{equation}
\mathcal{L}_{vis}^{C}
=
\mathbb{E}_{x\in\mathcal{C}}
\|M_\psi(x^d)\|_1
\end{equation}
\end{itemize}
Combining the above objectives, we formulate the camouflage generator optimization as follows:
\begin{flalign}
M_\psi^{*}
&=
\arg\min_{M_\psi}
\left\{
\begin{aligned}
&\lambda_o\mathcal{L}_{opp}^{C}
+\lambda_m\mathcal{L}_{margin}^{C} \\
&+\lambda_{cl}\mathcal{L}_{clean}^{C}
+\lambda_r\mathcal{L}_{vis}^{C}
\end{aligned}
\right. && \label{eq:camouflage_opt} \\
\text{s.t.}\quad
\theta_l
&=
\arg\min_{\theta}
\mathbb{E}_{\mathcal{D}^{*}}
\mathcal{L}(f_{\theta}(x),y) && \notag
\end{flalign}
where $\lambda_o$, $\lambda_m$, $\lambda_{cl}$, and $\lambda_r$ are balancing coefficients.

\subsection{Bilevel Optimization for Dual-Generator Learning}
In clean-label unlearning-activated backdoor attacks, trigger effectiveness depends on both the latent model after training and the post-unlearning model. Unlike conventional attacks optimized only for the trained model, the backdoor must remain dormant before unlearning and activate after camouflage samples are unlearned. We therefore simulate both latent-model training and subsequent unlearning during generator optimization.
Accordingly, we formulate dual-generator learning as a bilevel optimization problem. The outer level optimizes the trigger generator \(G_{\phi}\) and camouflage generator \(M_{\psi}\), while the inner level constructs the current training set, trains a surrogate latent model, and performs unlearning. The resulting latent and post-unlearning models provide feedback for updating both generators.
It is worth noting that, in real-world attack settings, attackers cannot access the deployed victim model or its training process. 
We therefore adopt a surrogate training strategy that uses a locally constructed shadow model closely approximating the victim architecture and attacker-accessible auxiliary data to simulate the training and unlearning processes.
The complete dual-generator optimization algorithm is
provided in Appendix A, Sec. D.

\section{Experiments}
\subsection{Experimental Settings}
\noindent\textbf{Datasets and Models.}
Following prior backdoor studies, we evaluate our method on two benchmark image classification datasets, CIFAR-10~\cite{krizhevsky2009learning} and ImageNet-10 that is constructed by randomly selecting 10 classes from the standard ImageNet dataset~\cite{deng2009imagenet}.
For the default evaluation, we use Pre-activation ResNet-18~\cite{he2016identity} on CIFAR-10 and ResNet-18~\cite{he2016deep} on ImageNet-10.
We further evaluate cross-architecture transfer with VGG-16~\cite{simonyan2014very} and MobileNetV2~\cite{sandler2018mobilenetv2}.
Both the trigger generator $G_\phi$ and camouflage generator $M_\psi$ use a U-Net~\cite{ronneberger2015u} backbone throughout all experiments.

\begin{table*}[t]
\centering
\setlength{\tabcolsep}{3.2pt}
\renewcommand{\arraystretch}{1.05}
\footnotesize
\begin{tabular}{cccccccccc}
\toprule
\multirow{3}{*}{Attack Method}
& \multirow{3}{*}{Unlearning Algorithm}
& \multicolumn{4}{c}{CIFAR-10}
& \multicolumn{4}{c}{ImageNet-10} \\
\cmidrule(lr){3-6} \cmidrule(lr){7-10}
&
& \multicolumn{2}{c}{Before Unlearn}
& \multicolumn{2}{c}{After Unlearn}
& \multicolumn{2}{c}{Before Unlearn}
& \multicolumn{2}{c}{After Unlearn} \\
\cmidrule(lr){3-4} \cmidrule(lr){5-6}
\cmidrule(lr){7-8} \cmidrule(lr){9-10}
&
& ASR & BA & ASR-U($\Delta$) & BA-U
& ASR & BA & ASR-U($\Delta$) & BA-U \\
\midrule

Ours & \multirow{5}{*}{SISA}
& \textbf{8.91} & 91.39 & 89.01(\textbf{+80.10}) & 90.82
& \textbf{10.67} & 80.34 & 80.30(\textbf{+69.63}) & 80.08 \\
UBA-Inf &
& 14.69 & 92.24 & 89.32(+74.63) & 91.04
& 17.22 & 81.50 & 84.81(+67.59) & 79.08 \\
UNCLEAN-Origin &
& 14.28 & 90.01 & 83.49(+69.21) & 89.84
& 30.25 & 78.42 & 65.67(+35.42) & 76.92 \\
UNCLEAN-Align &
& 15.78 & 88.57 & 76.73(+60.95) & 88.69
& 31.17 & 78.83 & 66.92(+35.75) & 77.75 \\

Sleeper Agent &
& 37.63 & 92.65 & 99.40(+61.77) & 91.47
& 36.23 & 81.07 & 85.83(+49.60) & 80.21 \\

\midrule

Ours & \multirow{5}{*}{PUMA}
& \textbf{13.74} & 93.38 & 86.99(\textbf{+73.25}) & 80.77
& \textbf{13.80} & 83.58 & 80.74(\textbf{+66.94}) & 70.08 \\
UBA-Inf &
& 17.36 & 90.39 & 84.16(+66.80) & 82.94
& 26.57 & 80.67 & 84.53(+57.96) & 66.17 \\
UNCLEAN-Origin &
& 22.84 & 87.95 & 47.43(+24.59) & 81.59
& 24.83 & 85.50 & 66.17(+41.34) & 68.75 \\
UNCLEAN-Align &
& 21.28 & 88.84 & 52.29(+31.01) & 79.39
& 23.75 & 84.83 & 62.00(+38.25) & 70.75 \\

Sleeper Agent &
& 42.81 & 93.96 & 67.63(+24.82) & 80.15
& 40.17 & 82.96 & 75.83(+35.66) & 72.87 \\

\midrule

Ours & \multirow{5}{*}{First-Order}
& \textbf{12.67} & 93.35 & 87.87(\textbf{+75.20}) & 83.24
& \textbf{14.35} & 83.21 & 82.13(\textbf{+67.78}) & 76.50 \\
UBA-Inf &
& 17.36 & 90.39 & 54.09(+36.73) & 84.59
& 26.57 & 80.67 & 52.04(+25.47) & 73.58 \\
UNCLEAN-Origin &
& 22.84 & 87.95 & 41.65(+18.81) & 70.66
& 24.83 & 85.50 & 50.25(+25.42) & 71.92 \\
UNCLEAN-Align &
& 21.28 & 88.84 & 39.45(+18.17) & 72.18
& 23.75 & 84.83 & 55.42(+31.67) & 73.08 \\

Sleeper Agent &
& 42.81 & 93.96 & 62.46(+19.65) & 81.54
& 40.17 & 82.96 & 72.50(+32.33) & 75.33 \\

\bottomrule
\end{tabular}
\caption{Main results and comparison with existing attacks across unlearning algorithms on CIFAR-10 and ImageNet-10.}
\label{tab:comparison_all}
\end{table*}
\noindent\textbf{Parameter and Attack Settings.}
Victim classifiers are trained for 250 epochs using SGD with momentum 0.9 and weight decay $5\times10^{-4}$. For CIFAR-10 and ImageNet-10, the batch sizes are 128 and 32, and the initial learning rates are 0.01 and 0.001, respectively. Class 0 is used as the target label. Before alternating optimization, an initial trigger generator is trained to construct the trigger-only reference model $f_{\theta_p}$ and initialize $G_\phi$. Both $G_\phi$ and $M_\psi$ are then optimized with an initial learning rate of 0.01. We set the poisoning and camouflage sample rates to $\rho_p=1\%$ and $\rho_c=3\%$, with perturbation strengths $\eta_G=\eta_C=0.15$. The loss coefficients are $\lambda_l=1$, $\lambda_a=2$, $\lambda_v=0.02$, and $\lambda_c=0.8$ for $G_\phi$, and $\lambda_o=2.5$, $\lambda_m=0.5$, $\lambda_{cl}=1.0$, and $\lambda_r=0.05$ for $M_\psi$. The margin hyperparameter is set to $\gamma=5$. The unlearning set $\mathcal{U}\subset\mathcal{C}'$ contains 250 samples for CIFAR-10 and 150 for ImageNet-10. We use $\tau=0.025$ for First-Order and PUMA and two shards for SISA. Unless otherwise specified, experiments are conducted on CIFAR-10 using First-Order unlearning for evaluation and all reported results are averaged over three independent runs with different random seeds. To further assess the visual consistency of the crafted samples, we provide visual examples in Appendix A, Sec. F.

\noindent \textbf{Unlearning Algorithms.}
To account for practical unlearning scenarios, we evaluate our attack under three representative unlearning algorithms that cover different design principles: SISA, First-Order, and PUMA, with further details provided in Appendix A, Sec. B.



\noindent \textbf{Evaluation Metrics.}
To evaluate attack performance, we report four metrics: Attack Success Rate (ASR), ASR after Unlearning (ASR-U), Benign Accuracy (BA), and Benign Accuracy after Unlearning (BA-U). 
ASR and ASR-U measure the proportion of non-target triggered inputs classified as the target class before and after unlearning respectively. 
BA and BA-U measure the clean classification accuracy before and after unlearning. 
An effective unlearning-activated backdoor should maintain low ASR before unlearning, high ASR-U after unlearning, and acceptable BA-U.

\subsection{Main Results and Comparisons}

Table~\ref{tab:comparison_all} reports the results on
CIFAR-10 and ImageNet-10 under different machine
unlearning algorithms. Our method consistently exhibits the
intended dormant-to-activated behavior, maintaining low ASR
before unlearning and substantially increasing ASR-U after
the requested camouflage samples are unlearned. 
Although BA-U decreases in some settings, the controlled experiments in Appendix A, Sec. E show that comparable degradation also occurs under equal-budget clean-sample removal, indicating that the observed activation is not merely caused by general model degradation.
We further compare our method with UBA-Inf~\cite{huang2024uba}, UNCLEAN~\cite{unclean}, and
Sleeper Agent~\cite{souri2022sleeper}. Specifically, UNCLEAN-Origin
follows its original clean-counterpart forgetting setting,
while UNCLEAN-Align uses the same forgetting budget on
samples that actually participate in training. For Sleeper
Agent, we construct naive removable camouflage samples by
applying its trigger construction to selected samples and
adding uniform random noise. As shown in
Table~\ref{tab:comparison_all}, our method consistently
achieves the lowest pre-unlearning ASR and the largest
$\Delta$ASR across datasets and unlearning algorithms.
Although UBA-Inf and Sleeper Agent sometimes achieve
higher absolute ASR-U, their higher pre-unlearning ASR
indicates premature backdoor exposure. UNCLEAN generally
produces a weaker ASR increase after unlearning. These
results demonstrate that the proposed removable camouflage
design more effectively coordinates pre-unlearning dormancy
and unlearning-triggered activation.

\begin{table}[t]
\centering
\setlength{\tabcolsep}{4pt}
\renewcommand{\arraystretch}{0.95}
\footnotesize
\begin{tabular}{lcccc}
\toprule
$|\mathcal{P}|:|\mathcal{C}|$ 
& ASR
& ASR-U
& BA
& BA-U \\
\midrule
1:0.5 & 35.08 & 79.64 & 93.44 & 86.23 \\
1:1   & 30.77 & 83.07 & 93.16 & 84.66 \\
1:2   & 21.84 & 85.62 & 93.28 & 83.91 \\
1:3   & 12.67 & 87.87 & 93.35 & 83.24 \\
\bottomrule
\end{tabular}

\caption{Effect of backdoor-to-camouflage sample allocation on CIFAR-10.}
\label{tab:ratio}
\end{table}
\subsection{Robustness and Sensitivity Analysis}
We analyze the robustness of the proposed attack under different attack configurations and deployment conditions. 

\textbf{Effect of poisoning and camouflage sample rates.}
We fix the backdoor sample rate at $\rho_p=1\%$ and vary the ratio between backdoor and camouflage samples.
As shown in Table~\ref{tab:ratio}, increasing camouflage proportion gradually lowers ASR before unlearning and raises ASR-U after unlearning.
This suggests that more camouflage samples provide stronger removable suppression, while BA remains relatively stable and BA-U decreases moderately after unlearning.

\begin{figure}[t]
    \centering
    \includegraphics[width=0.9\columnwidth]{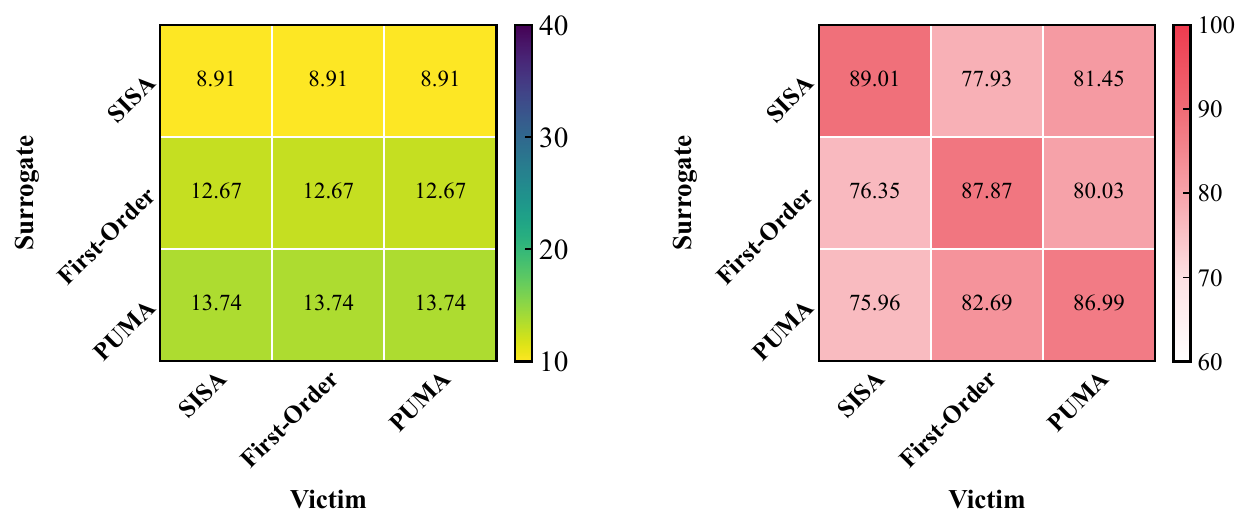}
   \caption{ASR before unlearning (left) and ASR-U after unlearning (right) across surrogate and victim algorithms.}
    \label{fig:cross_unlearning_transfer}
\end{figure}

\textbf{Transferability across Unlearning Algorithms.}
We further evaluate transferability across unlearning algorithms. 
As shown in
Fig.~\ref{fig:cross_unlearning_transfer}, matched settings
generally achieve the highest ASR-U, while most mismatched
settings retain effective activation and consistently low
pre-unlearning ASR. These results show that the learned
suppression--release relation transfers across unlearning
procedures, preserving dormancy before unlearning and
activation afterward.

\begin{figure}[t]
    \centering
    \includegraphics[width=0.9\columnwidth]{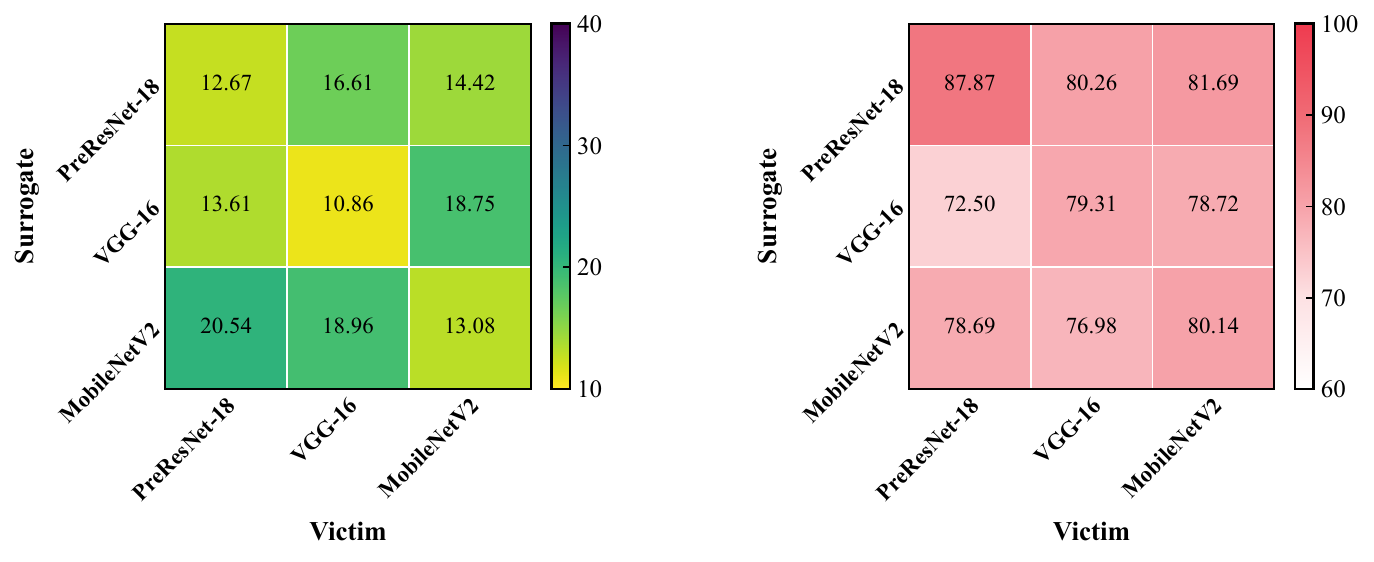}
   \caption{ASR before unlearning (left) and ASR-U after unlearning (right) across surrogate and victim models.}
    \label{fig:model_transferability}
\end{figure}

\textbf{Transferability across Model Architectures.}
We further examine whether the learned attack remains effective when the surrogate and victim architectures differ.
As shown in Fig.~\ref{fig:model_transferability}, the attack maintains low pre-unlearning ASR and high post-unlearning ASR-U under both matched and mismatched settings.
This indicates that the learned latent association and removable suppression are not specific to a particular architecture, supporting transfer to unseen victim models.

\subsection{Defense Evaluation}
We evaluate the proposed backdoor attack against fine-pruning~\cite{liu2018finepruning}, a representative defense that prunes neurons with low activation on clean samples.
Experiments are conducted on CIFAR-10 and ImageNet-10 under First-Order unlearning.
We further evaluate STRIP~\cite{gao2019strip} as a mainstream defense on both datasets, with detailed results provided in Appendix A, Sec. G.
As shown in Fig.~\ref{fig:defense_finepruning}, for latent models before unlearning, ASR remains low as neurons are gradually pruned, while BA stays high until aggressive pruning is applied.
For activated models after unlearning, ASR remains high across a wide range of pruning and drops clearly only when a large number of neurons are pruned.
However, this reduction is accompanied by a substantial decrease in BA, indicating that fine-pruning cannot remove the activated backdoor without significantly damaging benign model performance.
\begin{figure}[t]
    \centering
    \setlength{\tabcolsep}{1pt}
    \renewcommand{\arraystretch}{0.85}

    \begin{tabular}{cc}
        \includegraphics[width=0.49\linewidth]{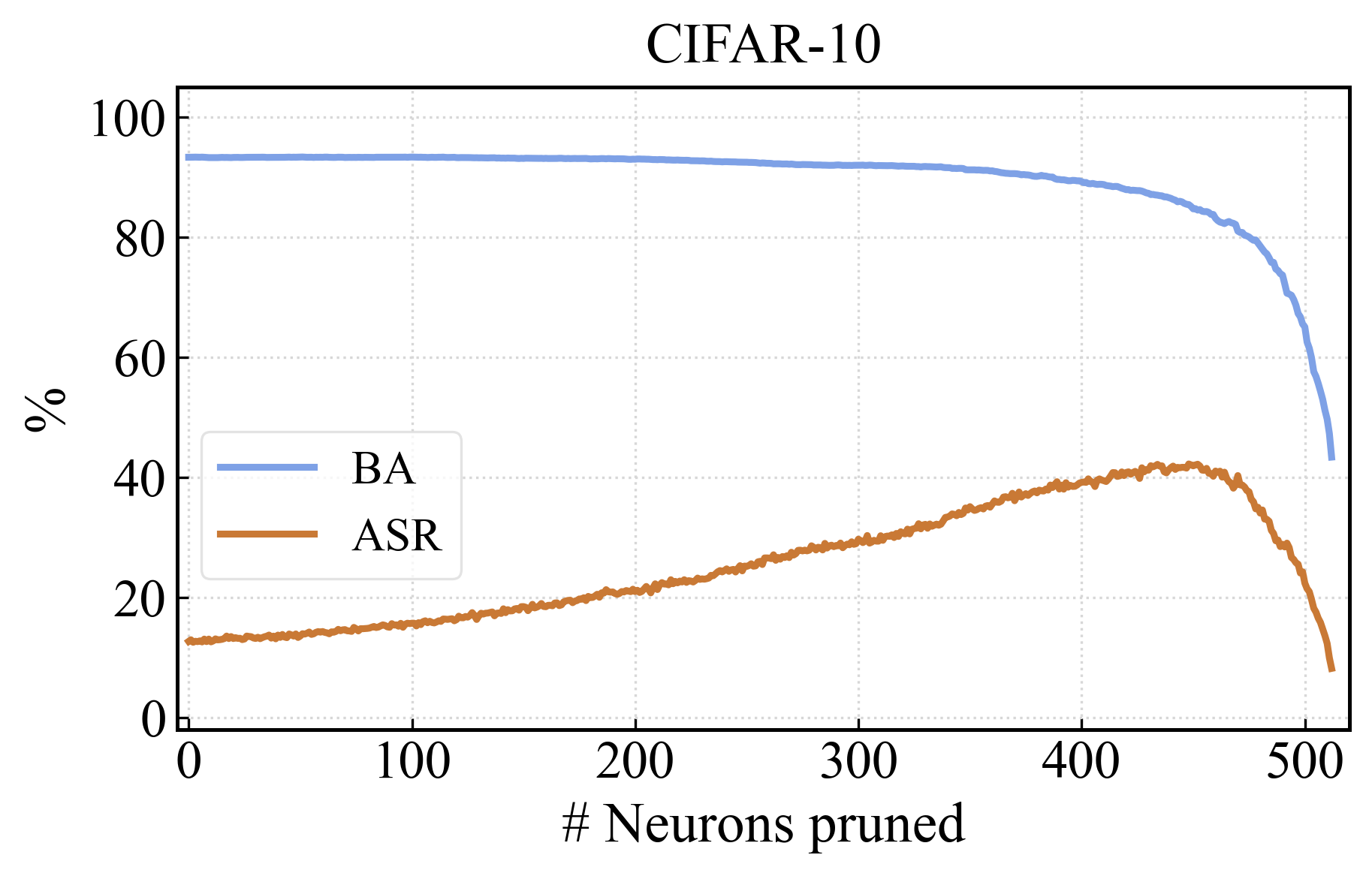} &
        \includegraphics[width=0.49\linewidth]{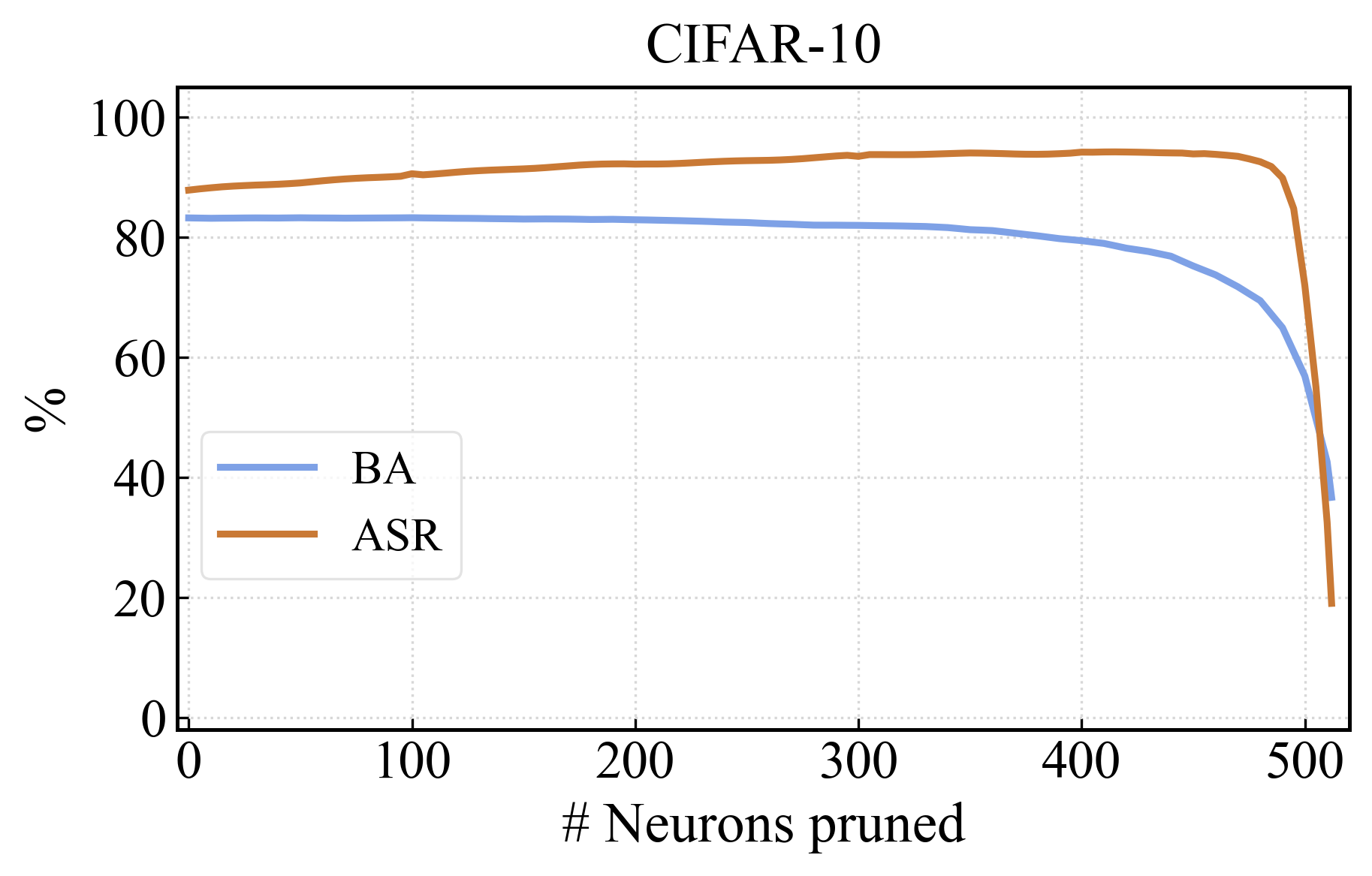} \\
        \small (a) & \small (b) \\[1mm]

        \includegraphics[width=0.49\linewidth]{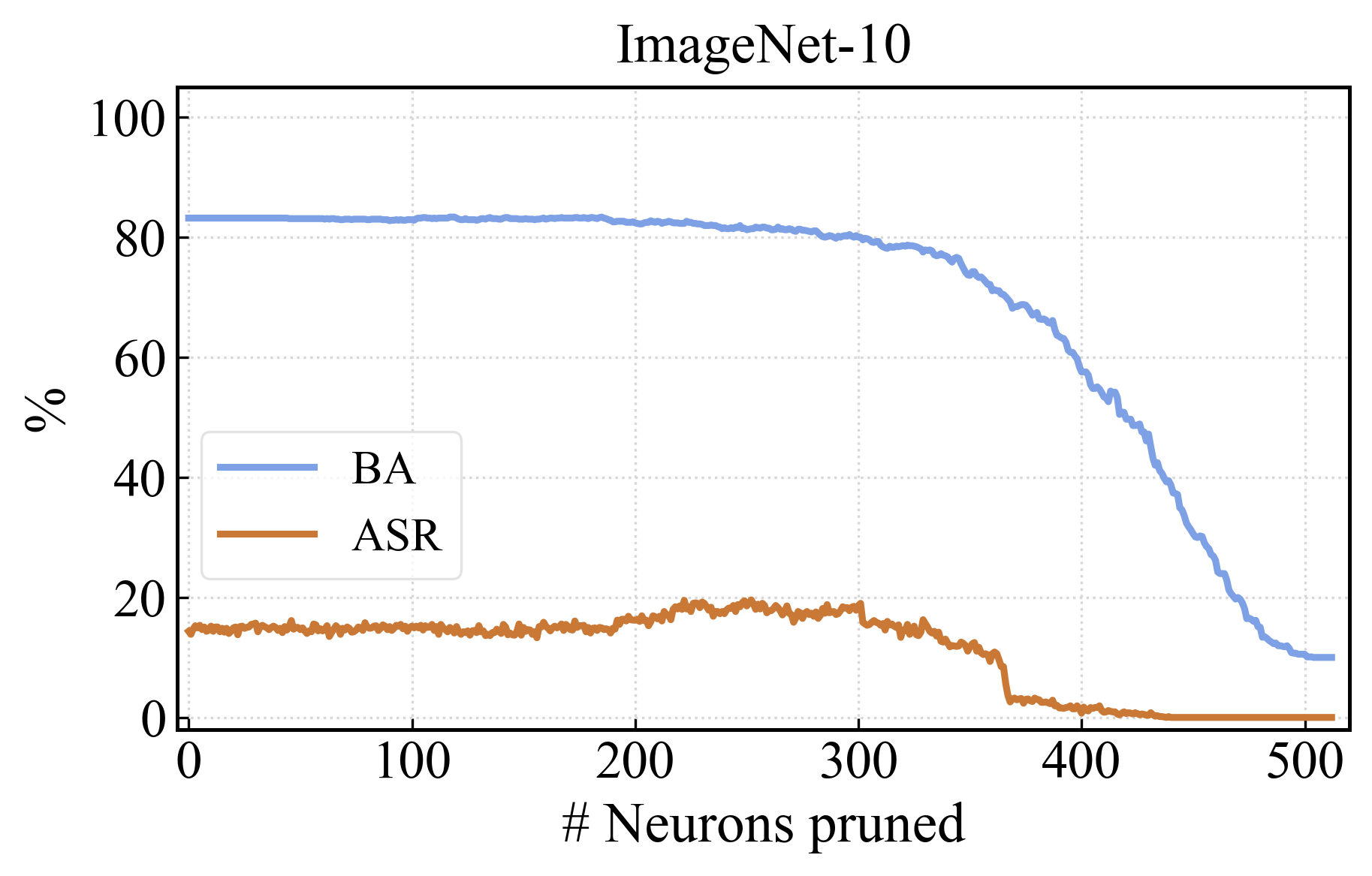} &
        \includegraphics[width=0.49\linewidth]{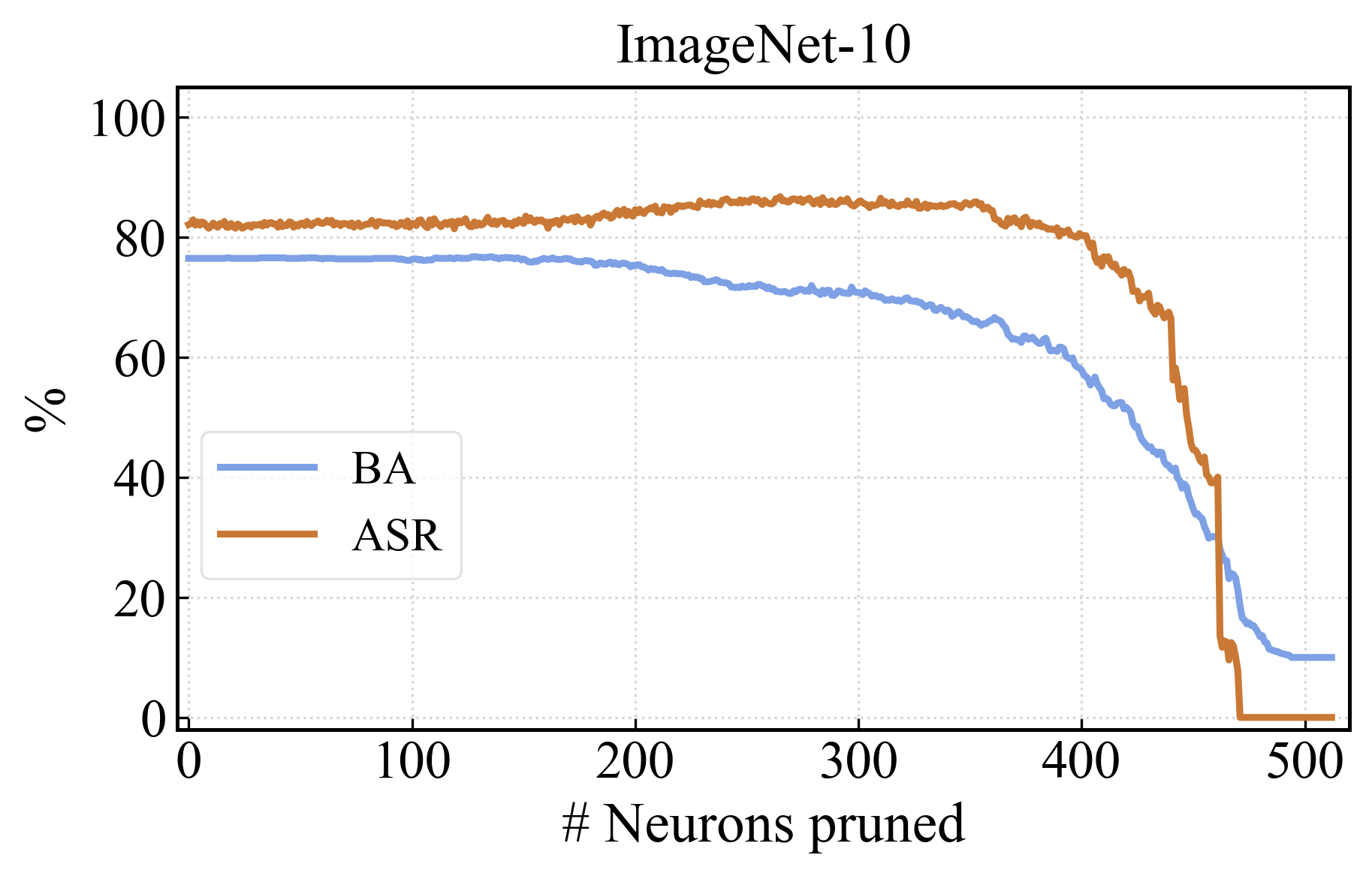} \\
        \small (c) & \small (d) 
    \end{tabular}

    \caption{Fine-pruning under First-Order unlearning:
(a, c) latent models and (b, d) activated models.}
    \label{fig:defense_finepruning}
\end{figure}
\begin{table}[t]
\centering
\setlength{\tabcolsep}{3pt}
\renewcommand{\arraystretch}{1.12}
\footnotesize
\begin{tabular}{lcccc}
\toprule
\multirow{2}{*}{Setting}
& \multicolumn{2}{c}{Before Unlearning}
& \multicolumn{2}{c}{After Unlearning} \\
\cmidrule(lr){2-3} \cmidrule(lr){4-5}
& ASR
& BA
& ASR-U ($\Delta$)
& BA-U \\
\midrule
\textbf{Full Method}
& \textbf{12.67} & \textbf{93.35} & \textbf{87.87 (+75.20)} & \textbf{83.24} \\

w/o $M_\psi$ 
& 61.74 & 93.20 & 69.41 (+7.67) & 92.50 \\

w/o $\mathcal{L}_{act}^{G}$ 
& 23.48 & 93.30 & 73.76 (+50.28) & 86.98 \\



w/o $\mathcal{L}_{opp}^{C}$ \& $\mathcal{L}_{margin}^{C}$ 
& 25.48 & 93.26 & 63.76 (+38.28) & 84.34 \\
\bottomrule
\end{tabular}
\caption{Ablation of key components on CIFAR-10.}
\label{tab:ablation}
\end{table}


\subsection{Ablation Study}
To assess the contribution of each component, we conduct ablation studies on CIFAR-10.
As shown in Table~\ref{tab:ablation}, removing the camouflage generator $M_\psi$ leads to a higher pre-unlearning ASR and a much smaller $\Delta$, indicating that removable camouflage samples are critical for maintaining the dormant state.
Removing the activation loss $\mathcal{L}_{act}^{G}$ results in a lower post-unlearning ASR-U, indicating that this objective is important for strengthening backdoor activation after unlearning.
Jointly removing $\mathcal{L}^{C}_{opp}$ and
$\mathcal{L}^{C}_{margin}$ increases the pre-unlearning ASR
and decreases ASR-U, substantially weakening the intended
dormant-to-activated transition.
Overall, these results confirm that the coordinated design of trigger learning and removable camouflage suppression is essential for dormancy before unlearning and enabling activation after unlearning.
\section{Conclusion}
In this paper, we propose a clean-label unlearning-activated backdoor
framework that jointly learns persistent trigger associations
and removable camouflage suppression through dual-generator
bilevel optimization. Experiments on CIFAR-10 and ImageNet-10
demonstrate consistent dormancy before unlearning and strong
activation afterward across different unlearning algorithms.
These results expose the security risk of adversarially crafted
forgetting requests and motivate more robust unlearning defenses.
\bibliography{references}

@article{gu2019badnets,
  title={{BadNets}: Evaluating Backdooring Attacks on
         Deep Neural Networks},
  author={Gu, Tianyu and Liu, Kang and
          Dolan-Gavitt, Brendan and Garg, Siddharth},
  journal={IEEE Access},
  volume={7},
  pages={47230--47244},
  year={2019}
}

@inproceedings{li2021invisible,
  author    = {Li, Yuezun and Li, Yiming and Wu, Baoyuan and Li, Longkang and He, Ran and Lyu, Siwei},
  title     = {Invisible Backdoor Attack with Sample-Specific Triggers},
  booktitle = {Proceedings of the IEEE/CVF International Conference on Computer Vision},
  pages     = {16463--16472},
  year      = {2021}
}

@inproceedings{nguyen2021wanet,
  author    = {Tuan Anh Nguyen and Anh Tuan Tran},
  title     = {{WaNet} - Imperceptible Warping-based Backdoor Attack},
  booktitle = {International Conference on Learning Representations},
  year      = {2021}
}

@inproceedings{nguyen2020inputaware,
  author    = {Nguyen, Tuan Anh and Tran, Anh Tuan},
  title     = {Input-Aware Dynamic Backdoor Attack},
  booktitle = {Advances in Neural Information Processing Systems},
  volume    = {33},
  pages     = {3454--3464},
  year      = {2020}
}

@inproceedings{huynh2024combat,
  author    = {Huynh, Tri and Nguyen, Dinh and Pham, Tuan and Tran, Anh},
  title     = {COMBAT: Alternated Training for Effective Clean-Label Backdoor Attacks},
  booktitle = {Proceedings of the AAAI Conference on Artificial Intelligence},
  volume    = {38},
  number    = {3},
  pages     = {2436--2444},
  year      = {2024}
}

@incollection{otto2018regulation,
  author    = {Otto, Marta},
  title     = {Regulation (EU) 2016/679 on the Protection of Natural Persons with Regard to the Processing of Personal Data and on the Free Movement of Such Data (General Data Protection Regulation--GDPR)},
  booktitle = {International and European Labour Law},
  pages     = {958--981},
  year      = {2018},
  publisher = {Nomos Verlagsgesellschaft mbH {\&} Co. KG}
}

@inproceedings{souri2022sleeper,
  title={Sleeper Agent: Scalable Hidden Trigger Backdoors for Neural Networks Trained from Scratch},
  author={Souri, Hossein and Fowl, Liam and Chellappa, Rama and Goldblum, Micah and Goldstein, Tom},
  booktitle={Advances in Neural Information Processing Systems},
  volume={35},
  pages={19165--19178},
  year={2022}
}

@inproceedings{bourtoule2021machine,
  author    = {Bourtoule, Lucas and Chandrasekaran, Varun and Choquette-Choo, Christopher A. and Jia, Hengrui and Travers, Adelin and Zhang, Baiwu and Lie, David and Papernot, Nicolas},
  title     = {Machine Unlearning},
  booktitle = {2021 IEEE Symposium on Security and Privacy},
  pages     = {141--159},
  year      = {2021},
  publisher = {IEEE}
}

@inproceedings{warnecke2023machine,
  author    = {Warnecke, Alexander and Pirch, Lukas and Wressnegger, Christian and Rieck, Konrad},
  title     = {Machine Unlearning of Features and Labels},
  booktitle = {Proceedings 2023 Network and Distributed System Security Symposium},
  year      = {2023},
  publisher = {Internet Society}
}

@inproceedings{wu2022puma,
  author    = {Wu, Ga and Hashemi, Masoud and Srinivasa, Christopher},
  title     = {PUMA: Performance Unchanged Model Augmentation for Training Data Removal},
  booktitle = {Proceedings of the AAAI Conference on Artificial Intelligence},
  volume    = {36},
  number    = {8},
  pages     = {8675--8682},
  year      = {2022}
}

@article{zhang2023exploiting,
  author  = {Zhang, Peixin and Sun, Jing and Tan, Ming and Wang, Xiaoyang},
  title   = {Exploiting Machine Unlearning for Backdoor Attacks in Deep Learning System},
  journal = {arXiv preprint arXiv:2310.10659},
  year    = {2023}
}

@inproceedings{liu2024backdoor,
  author    = {Liu, Zihao and Wang, Tianhao and Huai, Mengdi and Miao, Chenglin},
  title     = {Backdoor Attacks via Machine Unlearning},
  booktitle = {Proceedings of the AAAI Conference on Artificial Intelligence},
  volume    = {38},
  number    = {13},
  pages     = {14115--14123},
  year      = {2024}
}

@inproceedings{huang2024uba,
  title={{UBA-Inf}: Unlearning Activated Backdoor Attack with
         Influence-Driven Camouflage},
  author={Huang, Zirui and Mao, Yunlong and Zhong, Sheng},
  booktitle={33rd USENIX Security Symposium
             (USENIX Security 24)},
  pages={4211--4228},
  year={2024}
}

@article{unclean,
  author  = {Arazzi, Marco and Nocera, Antonino and {Vinod P}},
  title   = {When Forgetting Triggers Backdoors: A Clean Unlearning Attack},
  journal = {arXiv preprint arXiv:2506.12522},
  year    = {2025}
}

@inproceedings{wang2019neuralcleanse,
  author    = {Wang, Bolun and Yao, Yuanshun and Shan, Shawn and Li, Huiying and Viswanath, Bimal and Zheng, Haitao and Zhao, Ben Y.},
  title     = {Neural Cleanse: Identifying and Mitigating Backdoor Attacks in Neural Networks},
  booktitle = {2019 IEEE Symposium on Security and Privacy},
  pages     = {707--723},
  year      = {2019},
  publisher = {IEEE}
}

@inproceedings{liu2018finepruning,
  author    = {Liu, Kang and Dolan-Gavitt, Brendan and Garg, Siddharth},
  title     = {Fine-Pruning: Defending Against Backdooring Attacks on Deep Neural Networks},
  booktitle = {International Symposium on Research in Attacks, Intrusions, and Defenses},
  pages     = {273--294},
  year      = {2018},
  publisher = {Springer}
}

@inproceedings{gao2019strip,
  author    = {Gao, Yansong and Xu, Chang and Wang, Derui and Chen, Shiping and Ranasinghe, Damith C. and Nepal, Surya},
  title     = {{STRIP}: A Defence Against Trojan Attacks on Deep Neural Networks},
  booktitle = {Proceedings of the 35th Annual Computer Security Applications Conference},
  pages     = {113--125},
  year      = {2019}
}

@article{ahmed1974discrete,
  author  = {Ahmed, Nasir and Natarajan, T. and Rao, K. R.},
  title   = {Discrete Cosine Transform},
  journal = {IEEE Transactions on Computers},
  volume  = {C-23},
  number  = {1},
  pages   = {90--93},
  year    = {1974}
}

@techreport{krizhevsky2009learning,
  title={Learning Multiple Layers of Features from Tiny Images},
  author={Krizhevsky, Alex and Hinton, Geoffrey},
  institution={University of Toronto},
  year={2009}
}

@inproceedings{deng2009imagenet,
  author    = {Deng, Jia and Dong, Wei and Socher, Richard and Li, Li-Jia and Li, Kai and Fei-Fei, Li},
  title     = {ImageNet: A Large-Scale Hierarchical Image Database},
  booktitle = {2009 IEEE Conference on Computer Vision and Pattern Recognition},
  pages     = {248--255},
  year      = {2009},
  publisher = {IEEE}
}

@inproceedings{he2016identity,
  author    = {He, Kaiming and Zhang, Xiangyu and Ren, Shaoqing and Sun, Jian},
  title     = {Identity Mappings in Deep Residual Networks},
  booktitle = {European Conference on Computer Vision},
  pages     = {630--645},
  year      = {2016},
  publisher = {Springer}
}

@inproceedings{he2016deep,
  author    = {He, Kaiming and Zhang, Xiangyu and Ren, Shaoqing and Sun, Jian},
  title     = {Deep Residual Learning for Image Recognition},
  booktitle = {Proceedings of the IEEE Conference on Computer Vision and Pattern Recognition},
  pages     = {770--778},
  year      = {2016}
}

@inproceedings{ronneberger2015u,
  author    = {Ronneberger, Olaf and Fischer, Philipp and Brox, Thomas},
  title     = {U-Net: Convolutional Networks for Biomedical Image Segmentation},
  booktitle = {International Conference on Medical Image Computing and Computer-Assisted Intervention},
  pages     = {234--241},
  year      = {2015},
  publisher = {Springer}
}

@inproceedings{simonyan2014very,
  title={Very Deep Convolutional Networks for Large-Scale Image Recognition},
  author={Simonyan, Karen and Zisserman, Andrew},
  booktitle={International Conference on Learning Representations},
  year={2015}
}

@inproceedings{sandler2018mobilenetv2,
  title={MobileNetV2: Inverted Residuals and Linear Bottlenecks},
  author={Sandler, Mark and Howard, Andrew and Zhu, Menglong and Zhmoginov, Andrey and Chen, Liang-Chieh},
  booktitle={Proceedings of the IEEE Conference on Computer Vision and Pattern Recognition},
  pages={4510--4520},
  year={2018}
}

@inproceedings{alam2025reveil,
  title={{ReVeil}: Unconstrained Concealed Backdoor Attack on Deep Neural Networks Using Machine Unlearning},
  author={Alam, Manaar and Lamri, Hithem and Maniatakos, Michail},
  booktitle={2025 62nd ACM/IEEE Design Automation Conference (DAC)},
  pages={1--7},
  year={2025},
  organization={IEEE}
}

@inproceedings{nguyen2025wicked,
  title={Wicked Oddities: Selectively Poisoning for Effective Clean-Label Backdoor Attacks},
  author={Nguyen, Quang H. and Nguyen, Ngoc-Hieu and Ta, The-Anh and Nguyen-Tang, Thanh and Wong, Kok-Seng and Hoang, Thanh-Tung and Doan, Khoa D.},
  booktitle={International Conference on Learning Representations},
  year={2025}
}

\clearpage
\newcommand{\appsubsection}[1]{%
  \par\medskip
  \noindent\textit{#1}\par
  \nobreak\smallskip
}




\begin{center}
\textbf{\MakeUppercase{Appendix A}}\\
\textbf{\MakeUppercase{Additional Details and Experimental Evaluations}}
\end{center}
\par\smallskip

This appendix provides further technical details and supplementary analyses to facilitate reproducibility and a deeper
understanding of our experimental methodology.
\newlength{\sampleimagesize}
\setlength{\sampleimagesize}{0.17\textwidth}

\newcommand{\sampleimage}[1]{%
  \includegraphics[
    width=\sampleimagesize,
    height=\sampleimagesize
  ]{#1}%
}
\appsubsection{A. Datasets and Preprocessing}

In all of our comparative experiments, we evaluated every
method on the CIFAR-10 and ImageNet-10 datasets. For
fairness, all methods use identical data splits and preprocessing pipelines. CIFAR-10 consists of 50,000 training images and 10,000 test images, covering 10 object categories.
ImageNet-10 is constructed by randomly selecting 10 mutually exclusive classes from the standard ImageNet dataset. All images
are resized to 32 × 32 for CIFAR-10 and 224 × 224 for
ImageNet-10. For both datasets, we apply standard normalization using the per-channel mean and standard deviation.
Data augmentation includes random horizontal flipping, random rotation and random cropping for CIFAR-10, and random
resized cropping for ImageNet-10.
\appsubsection{B. Unlearning Methods Adopted}

In our experiments, we used three unlearning strategies: SISA, First-Order unlearning, and PUMA unlearning. These methods are designed to remove the influence of requested training samples while preserving the model's utility on the remaining data.
\begin{itemize}
    \item \textbf{SISA}: 
An exact unlearning framework that organizes training data via sharding and slicing, generates final results through sub-model aggregation, and handles unlearning by retraining only the affected shard.

    \item \textbf{First-Order}: 
A first-order approximate unlearning algorithm that applies gradient ascent to the requested forgetting samples to induce the forgetting effect.

\item \textbf{PUMA}: 
    An approximate unlearning algorithm that estimates the influence of removed samples and compensates for their removal by optimally reweighting the remaining data.
\end{itemize}
\smallskip
\appsubsection{C. Baseline Adaptation and Evaluation Settings}
To ensure a fair comparison, all methods are evaluated under a unified clean-label protocol, where all crafted samples retain their original labels and share identical experimental and machine-unlearning settings.
For UBA-Inf, we instantiate its influence-driven camouflage framework using the LC backdoor adopted in its original evaluation. For UNCLEAN, we report two variants. UNCLEAN-Origin follows the original clean-counterpart forgetting protocol, whereas UNCLEAN-Align restricts the removal set to records that participate in victim-model training and uses the same forgetting budget as our method.
Sleeper Agent does not originally include a removable suppression mechanism for unlearning-triggered activation. We therefore preserve its original clean-label trigger construction and augment it with a naive mitigation-sample construction. All mitigation samples preserve their original labels, and their number and forgetting budget are matched to the camouflage setting of our method. This adaptation is used to evaluate whether a conventional clean-label backdoor combined with non-optimized removable mitigation samples can realize a comparable dormant-to-activated transition.
\begin{algorithm}[t]
\caption{Bilevel Optimization for Dual-Generator Learning}
\label{alg:dual_generator}
\begin{algorithmic}[1]

\Statex \textbf{Input:} training set $\mathcal{D}$, backdoor subset $\mathcal{P}$,
camouflage subset $\mathcal{C}$, target label $y_t$,
generators $G_{\phi}$ and $M_{\psi}$,
reference models $f_{\theta_c}$ and $f_{\theta_p}$,
unlearning algorithm $\mathcal{F}_{\mathrm{unlearn}}$,
optimization rounds $R$

\Statex \textbf{Output:} Optimized generators
$G_{\phi^*}$ and $M_{\psi^*}$

\State Initialize generator parameters $\phi$ and $\psi$

\For{each outer-level round $r \in [1,R]$}

     \Statex \textit{// Outer-level sample construction induced by current generators}

    \State $\mathcal{P}_r \leftarrow
    \{(B_{\phi}(x),y)\mid(x,y)\in\mathcal{P}\}$

    \State $\mathcal{C}_r \leftarrow
    \{(C_{\phi,\psi}(x),y)\mid(x,y)\in\mathcal{C}\}$

    \State $\mathcal{D}_r \leftarrow
    (\mathcal{D}\setminus(\mathcal{P}\cup\mathcal{C}))
    \cup\mathcal{P}_r\cup\mathcal{C}_r$

    \Statex \textit{// Inner-level training and unlearning simulation}

    \State $\theta_l^r \leftarrow
    \mathrm{Train}(\theta_l^{r-1},\mathcal{D}_r)$

    \State Choose an unlearning set
    $\mathcal{U}_r\subseteq\mathcal{C}_r$

    \State $\theta_u^r \leftarrow
    \mathcal{F}_{\mathrm{unlearn}}
    (\theta_l^r,\mathcal{U}_r)$

   \Statex \textit{// Outer-level generator update with inner-level feedback}

    \State Compute the trigger-generator objective
    \Statex \hspace{\algorithmicindent}
    $\mathcal{L}_{G}^{r}
    =
    \lambda_l\mathcal{L}_{latent}^{G}
    +\lambda_a\mathcal{L}_{act}^{G}
    +\lambda_v\mathcal{L}_{vis}^{G}
    +\lambda_c\mathcal{L}_{clean}^{G}$

    \State Update the trigger generator
    \Statex \hspace{\algorithmicindent}
    \[
\phi \leftarrow \phi-\alpha_G\nabla_{\phi}\mathcal{L}_G^r
\]

    \State Compute the camouflage-generator objective
    \Statex \hspace{\algorithmicindent}
    $\mathcal{L}_{C}^{r}
    =
    \lambda_o\mathcal{L}_{opp}^{C}
    +\lambda_m\mathcal{L}_{margin}^{C}
    +\lambda_{cl}\mathcal{L}_{clean}^{C}
    +\lambda_r\mathcal{L}_{vis}^{C}$

    \State Update the camouflage generator
    \Statex \hspace{\algorithmicindent}
    \[
\psi \leftarrow \psi-\alpha_C\nabla_{\psi}\mathcal{L}_C^r
\]

\EndFor

\State \Return $G_{\phi^*}=G_{\phi}$ and
$M_{\psi^*}=M_{\psi}$

\end{algorithmic}
\end{algorithm}
\appsubsection{D. Bilevel Optimization for Dual-Generator Learning}

Algorithm~\ref{alg:dual_generator} presents the complete optimization procedure of the proposed dual-generator framework. At each outer-level round, the current trigger and camouflage generators construct the corresponding backdoor and camouflage samples, which are combined with the remaining clean data to form the surrogate training set. The inner level then updates the latent surrogate model and simulates machine unlearning on a selected subset of camouflage samples, yielding dynamically refreshed latent and post-unlearning model states. These model states provide state-specific feedback for the outer-level optimization and are treated as fixed when updating the generators, with gradients propagated through the generated samples rather than through the surrogate training or unlearning procedures. Accordingly, the trigger generator is optimized to preserve dormancy in the latent model while inducing target behavior after unlearning, whereas the camouflage generator learns a label-consistent suppressive influence that can be removed by the subsequent unlearning request. Repeating this alternating process coordinates persistent latent trigger associations with removable camouflage suppression.
\appsubsection{E. Verification of
Camouflage-Removal-Induced Activation}

To verify whether the post-unlearning activation is
attributable to the specific removal of camouflage samples,
rather than to general model-performance degradation caused
by the unlearning operation itself, we conduct a series of
verification experiments under the same experiment setting.
We first apply the unlearning operation to randomly selected
clean training samples from a clean model. This setting
quantifies the general benign-accuracy degradation caused by
equal-budget unlearning. We then evaluate a poisoned model
trained with trigger-bearing samples generated by the same
optimized trigger generator as the full method, but without
any camouflage samples. The same number of randomly
selected clean samples is unlearned from this model to
examine whether camouflage samples are necessary for
suppressing the trigger association and maintaining the
dormant state before unlearning.
Finally, for the same full latent model, we compare two
removal settings. In the first setting, the same unlearning
update is applied using the clean source counterparts of the
selected camouflage samples. In the second setting, the
actual camouflage samples $\mathcal{U}\subset\mathcal{C}'$are unlearned.
This comparison examines whether the strong activation is
specifically associated with removing the learned influence
of the camouflage samples, rather than merely with the
corresponding sample content or an equal-budget model
update.

As shown in
Table~\ref{tab:camouflage_removal_verification}, unlearning
random clean samples under the same configuration also
causes a comparable decrease in benign accuracy, indicating
that the observed BA-U reduction is not specific to
camouflage removal. Without camouflage samples, the
poisoned model already exhibits a high pre-unlearning ASR,
confirming their role in maintaining backdoor dormancy.
More importantly, for the same full latent model, removing
the actual camouflage samples produces substantially
stronger activation than the clean-source counterpart
control. Although BA-U decreases further, the additional
degradation remains modest relative to the substantial
increase in ASR. Overall, these results demonstrate that the camouflage
samples are essential for maintaining the dormant state and
that the subsequent dormant-to-activated transition is
primarily driven by removing their learned suppressive
influence.

\begin{table*}[t]
\centering
\small
\setlength{\tabcolsep}{5.5pt}

\begin{tabular}{llcccc}
\toprule
Training Configuration
& Removal Set
& ASR
& BA
& ASR-U ($\Delta$)
& BA-U \\
\midrule

Clean only
& Random clean samples
& 3.93
& 93.68
& 7.29 ({+}3.36)
& 86.29 \\

Backdoor only
& Random clean samples
& 60.72
& 92.49
& 83.64 ({+}22.92)
& 85.65 \\

Backdoor + camouflage
& Clean source counterparts
& 12.67
& 93.35
& 36.78 ({+}24.11)
& 86.84 \\

Backdoor + camouflage
& Camouflage samples $\mathcal{U}$
& 12.67
& 93.35
& \textbf{87.87 ({+}75.20)}
& 83.24 \\

\bottomrule
\end{tabular}

\caption{Verification experiments for
camouflage-removal-induced activation on CIFAR-10.}
\label{tab:camouflage_removal_verification}
\end{table*}
\appsubsection{F. Visual Examples of Crafted Samples}

\begin{figure*}[t]
\centering
\setlength{\tabcolsep}{4pt}
\setlength{\fboxsep}{0pt}
\setlength{\fboxrule}{0.3pt}
\renewcommand{\arraystretch}{1.0}

\begin{tabular}{cccc}


\sampleimage
{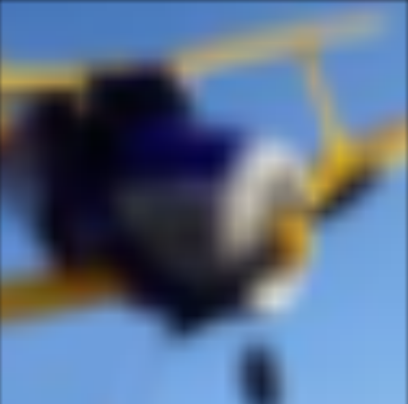}
&
\sampleimage
{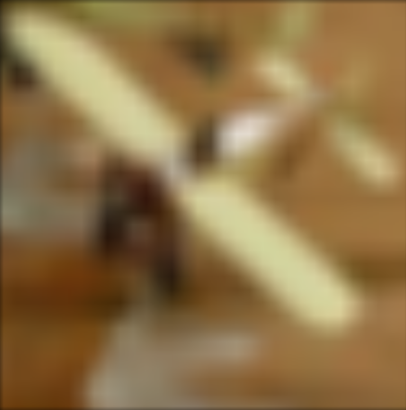}
&
\sampleimage
{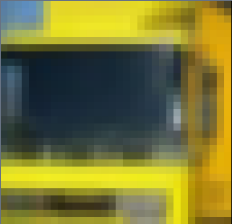}
&
\sampleimage
{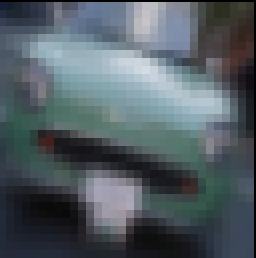}
\\[3mm]

\sampleimage
{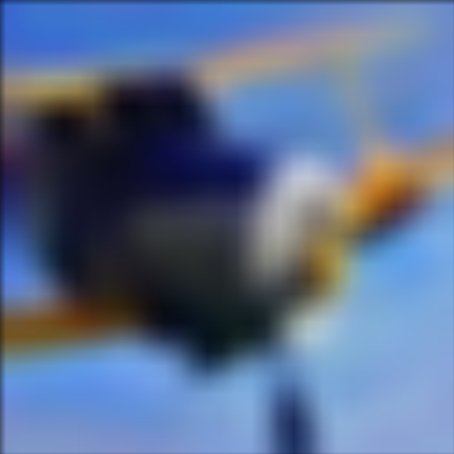}
&
\sampleimage
{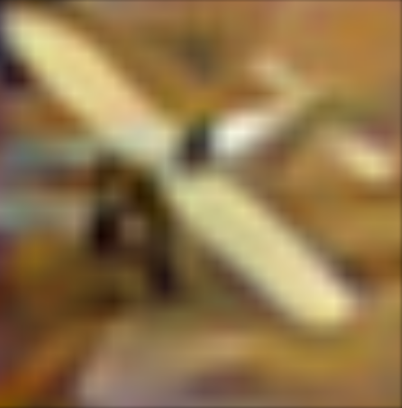}
&
\sampleimage
{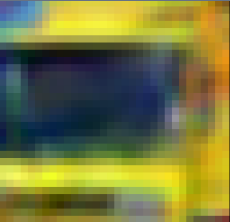}
&
\sampleimage
{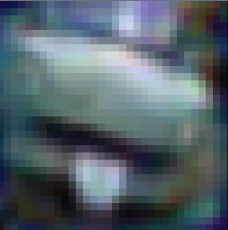}
\\[2mm]

\multicolumn{2}{c}{\textbf{(a) CIFAR-10 (Poisoned)}}
&
\multicolumn{2}{c}{\textbf{(b) CIFAR-10 (Camouflage)}}
\\[5mm]


\sampleimage
{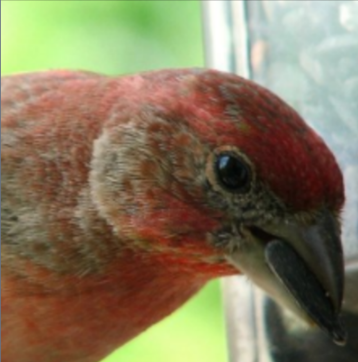}
&
\sampleimage
{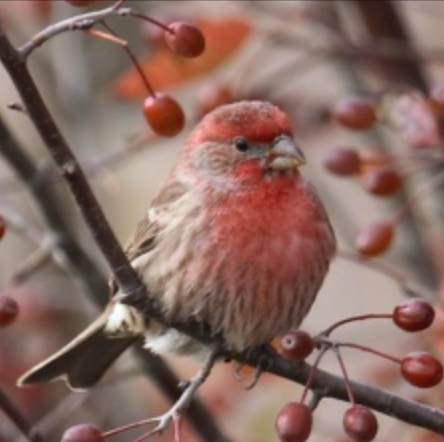}
&
\sampleimage
{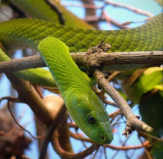}
&
\sampleimage
{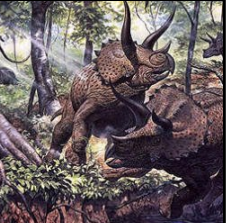}
\\[3mm]

\sampleimage
{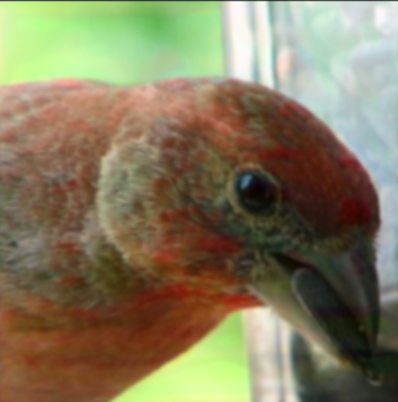}
&
\sampleimage
{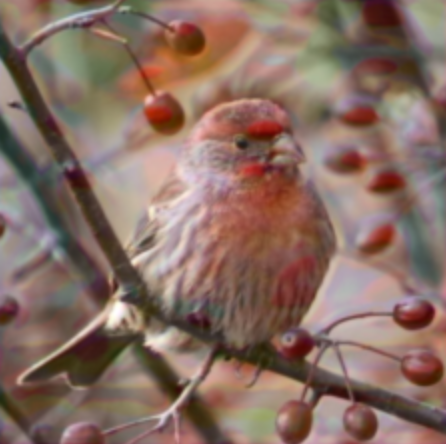}
&
\sampleimage
{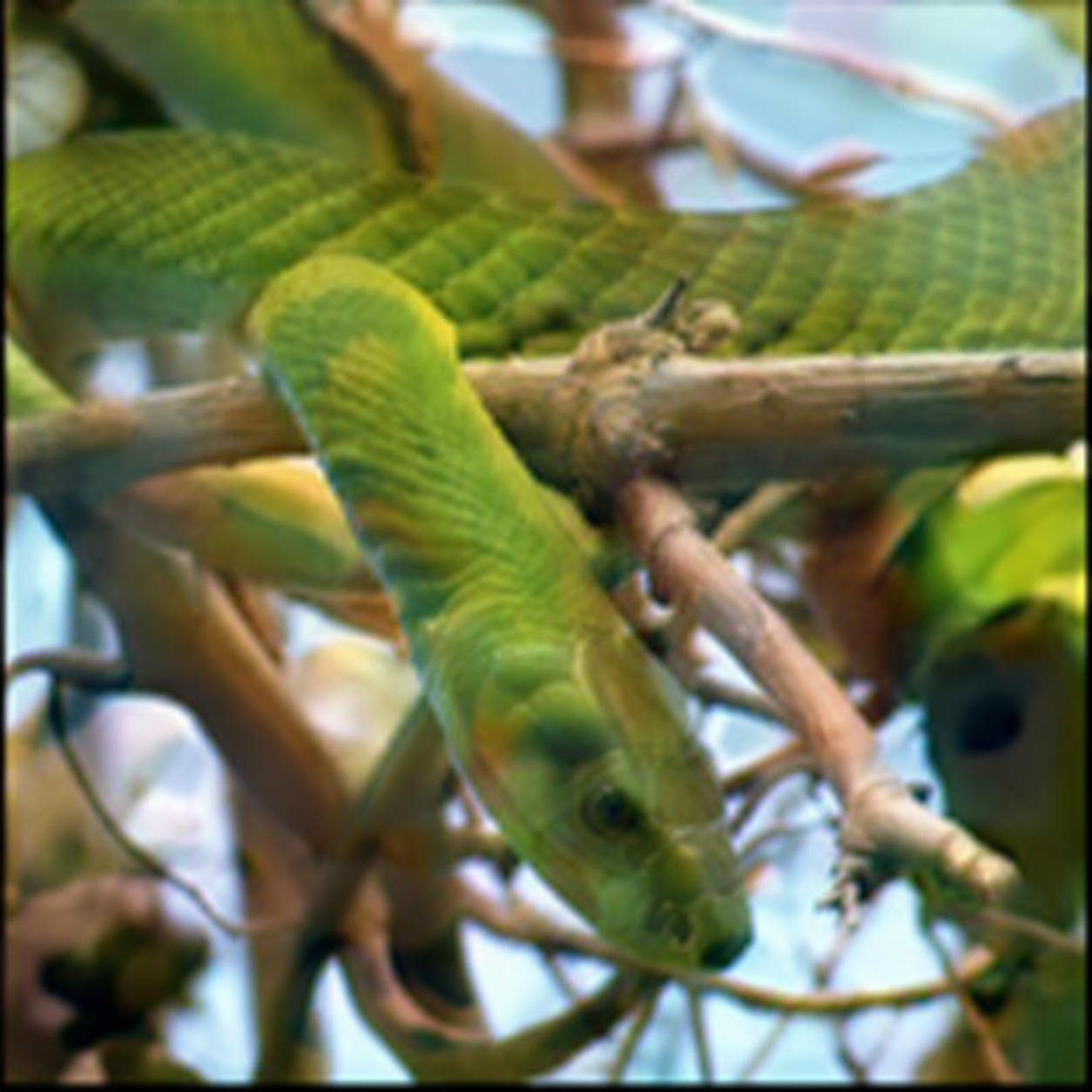}
&
\sampleimage
{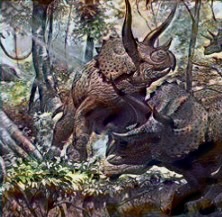}
\\[2mm]

\multicolumn{2}{c}{\textbf{(c) ImageNet-10 (Poisoned)}}
&
\multicolumn{2}{c}{\textbf{(d) ImageNet-10 (Camouflage)}}

\end{tabular}

\caption{Visual examples of crafted samples on CIFAR-10 and
ImageNet-10.}
\label{fig:visual_crafted_samples}
\end{figure*}

To qualitatively assess the visual consistency of the crafted
samples, we present poisoned and
camouflage samples from CIFAR-10 and ImageNet-10 together
with their corresponding clean source images. As shown in
Fig.~\ref{fig:visual_crafted_samples}, for each dataset, the
upper row contains the clean source images and the lower row
contains their corresponding crafted samples. The first two
columns show poisoned samples, while the last two
columns show camouflage samples.
The poisoned and camouflage samples preserve the primary
object appearance and recognizable semantic content of
their clean counterparts, without introducing conspicuous
localized patterns or altering the semantics associated with
their original labels. These examples illustrate that both
types of crafted samples maintain visual and label
consistency under the clean-label data-submission setting.

\appsubsection{G. Additional Defense Evaluation}

In this section, we further evaluate the resistance of the
proposed attack to STRIP, complementing the fine-pruning
evaluation reported in the main paper. Experiments are
conducted on CIFAR-10 and ImageNet-10 under First-Order
unlearning, considering both the latent model before unlearning
and the activated model after unlearning. STRIP is implemented
following its original evaluation protocol.

\begin{figure*}[t]
    \centering

    \begin{minipage}[t]{0.41\textwidth}
        \centering
        \includegraphics[width=0.95\linewidth]
        {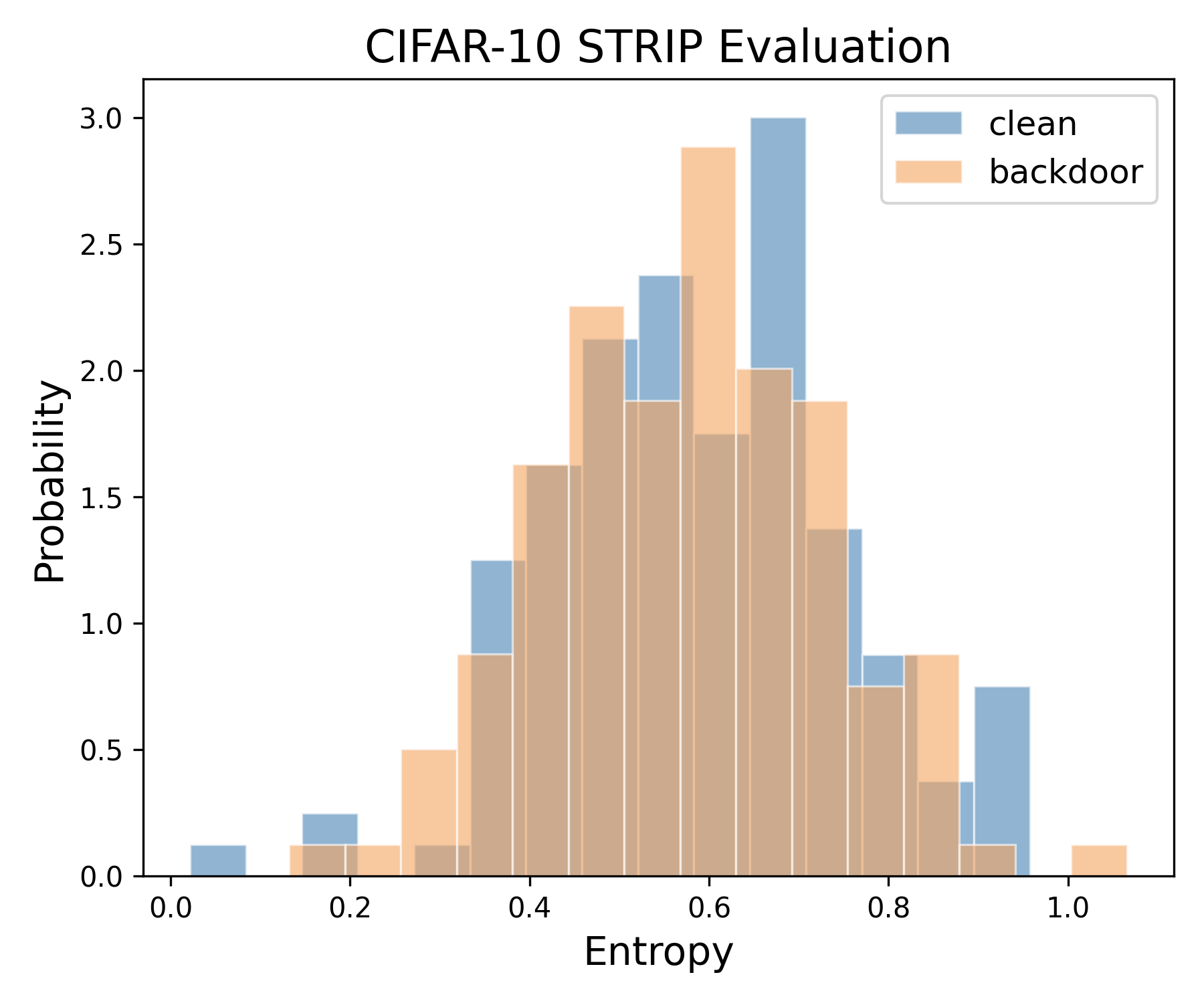}

        {\small (a) }
    \end{minipage}
    \hspace{0.04\textwidth}
    \begin{minipage}[t]{0.41\textwidth}
        \centering
        \includegraphics[width=0.95\linewidth]
        {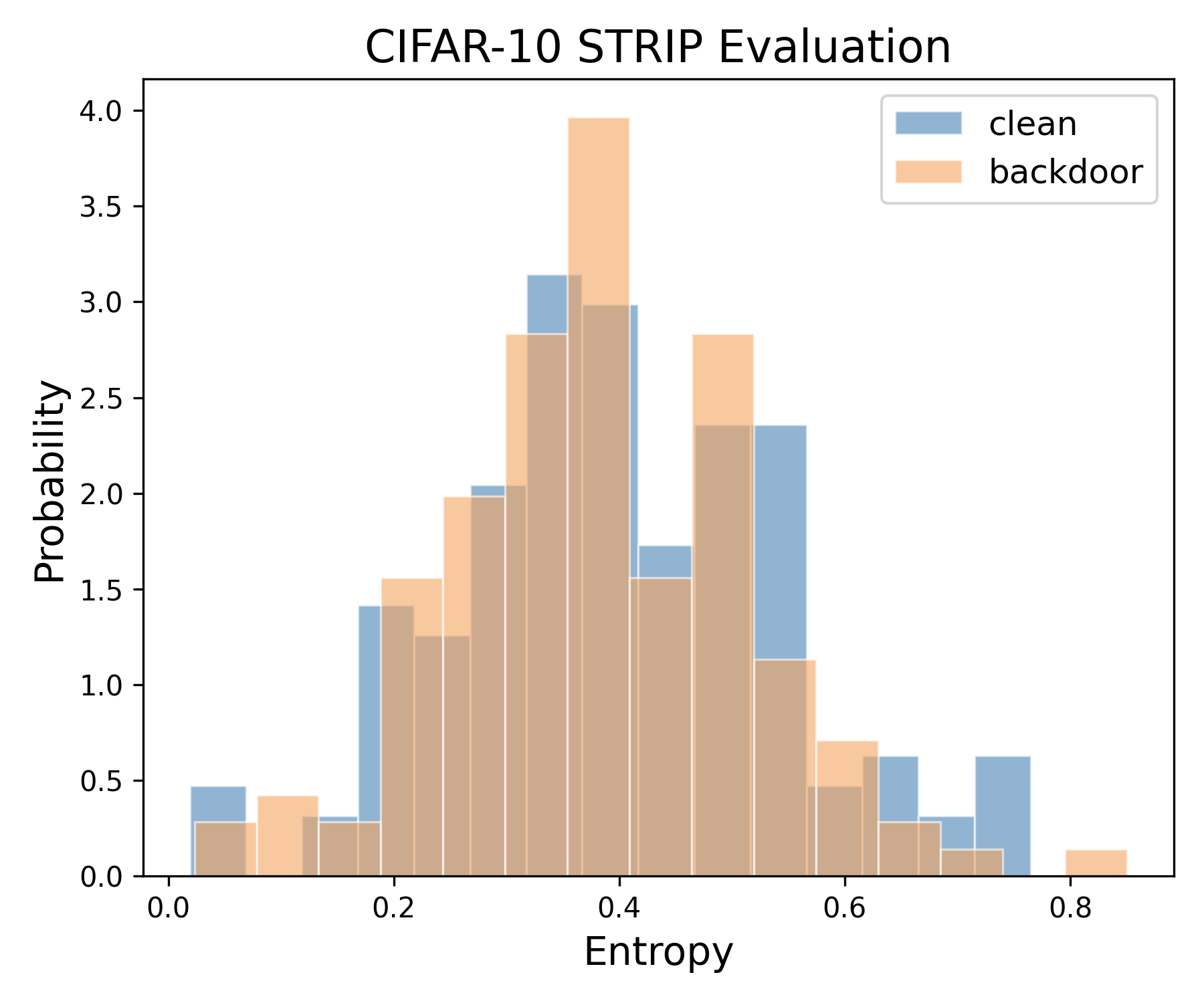}

        {\small (b)}
    \end{minipage}

    \vspace{1.5mm}

    \begin{minipage}[t]{0.41\textwidth}
        \centering
        \includegraphics[width=0.95\linewidth]
        {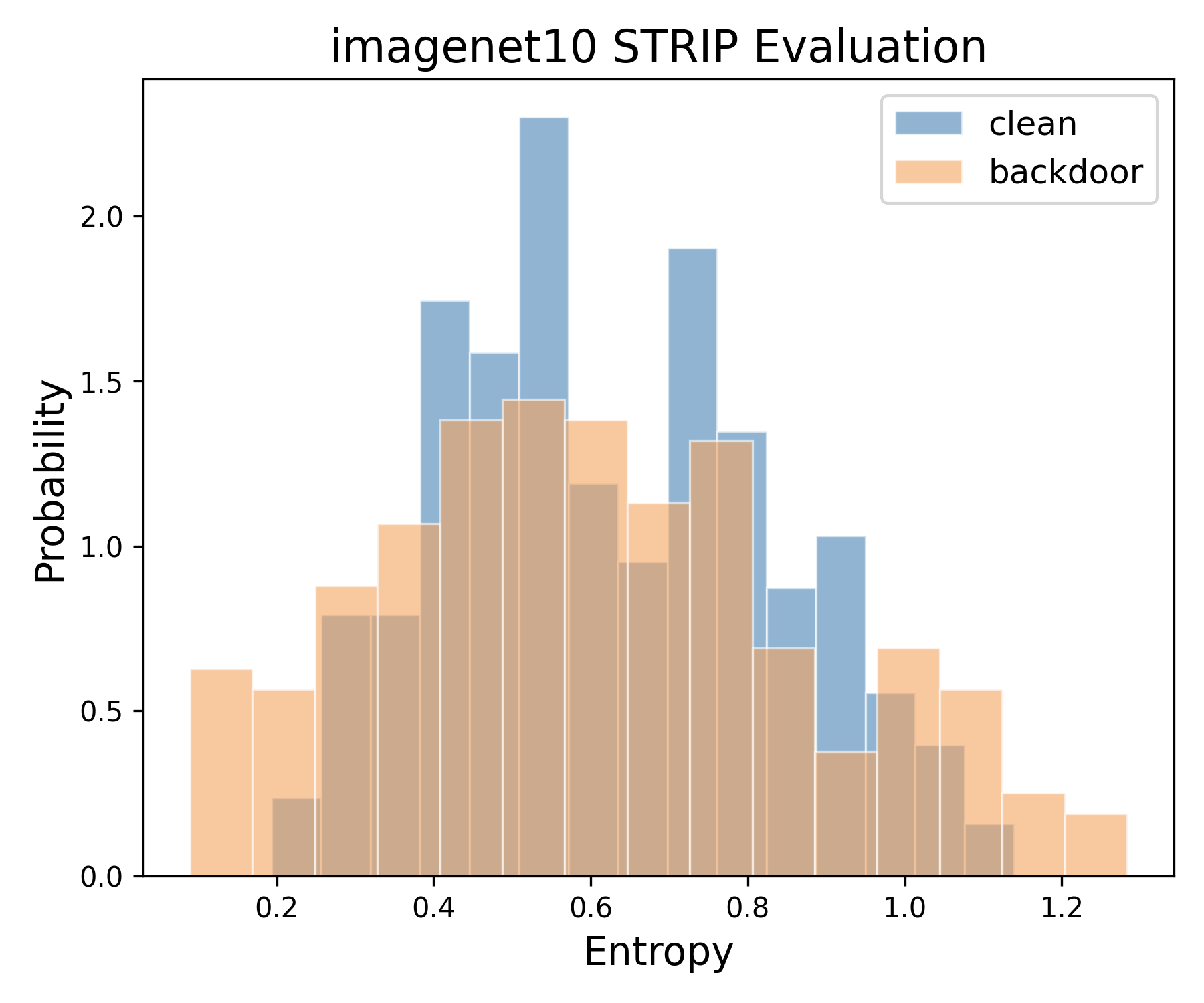}

        {\small (c)}
    \end{minipage}
    \hspace{0.04\textwidth}
    \begin{minipage}[t]{0.41\textwidth}
        \centering
        \includegraphics[width=0.95\linewidth]
        {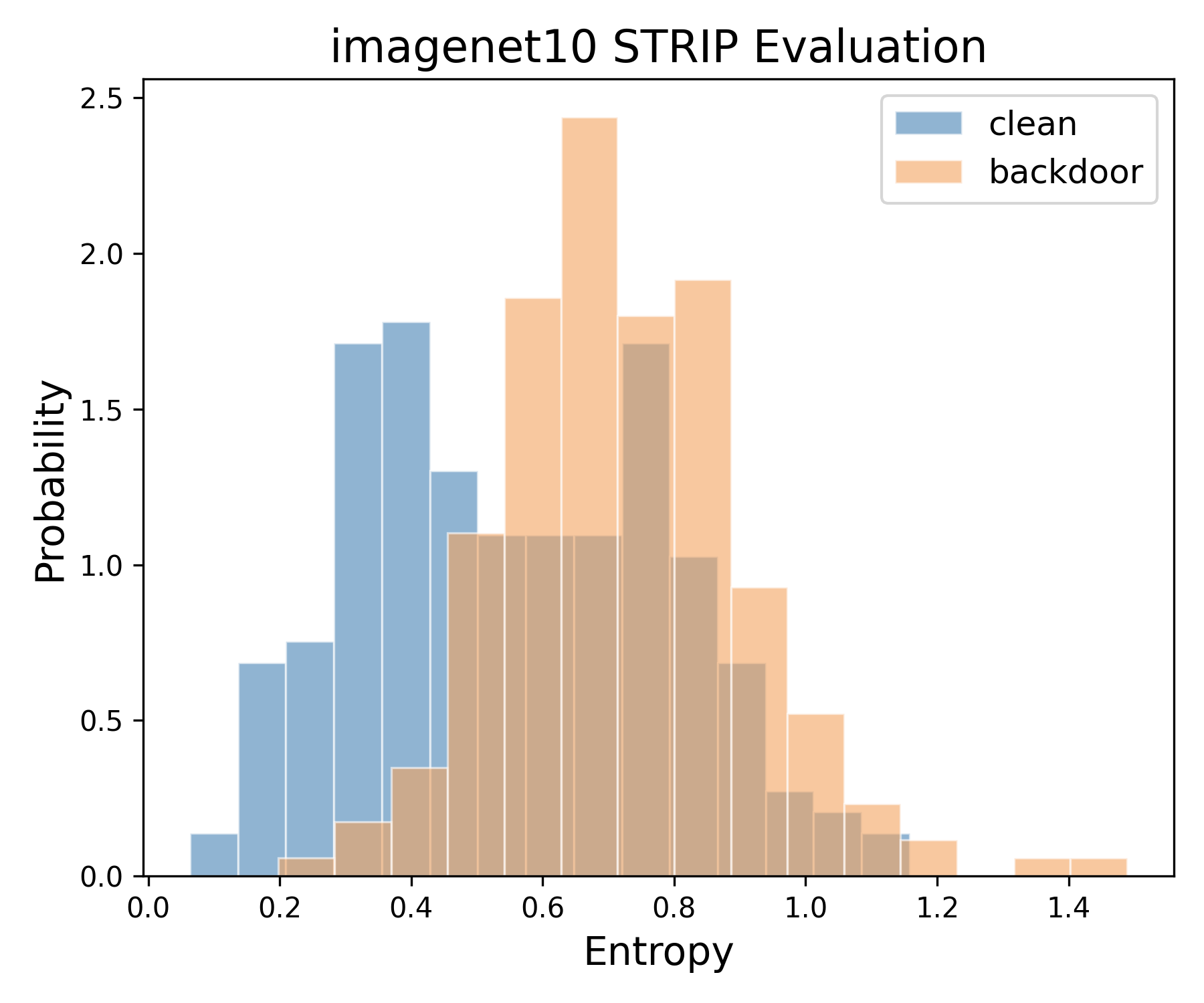}

        {\small (d)}
    \end{minipage}

    \caption{Entropy distributions of clean and trigger-bearing
    inputs under STRIP with First-Order unlearning:
    (a) CIFAR-10 latent model before unlearning,
    (b) CIFAR-10 activated model after unlearning,
    (c) ImageNet-10 latent model before unlearning, and
    (d) ImageNet-10 activated model after unlearning.
    }
    \label{fig:strip}
\end{figure*}

\smallskip
\noindent\textbf{STRIP.}
STRIP detects potential backdoor inputs by repeatedly
superimposing them with clean images and measuring the entropy
of the resulting predictions. Conventional backdoor inputs tend
to preserve the attacker-specified prediction under such
perturbations and therefore exhibit abnormally low entropy.
As shown in Fig.~\ref{fig:strip}, the entropy distributions of
clean and trigger-bearing inputs substantially overlap for the
latent models on both CIFAR-10 and ImageNet-10. After
unlearning, substantial overlap remains on CIFAR-10, while the
trigger-bearing inputs on ImageNet-10 shift toward higher rather
than lower entropy. Therefore, the proposed attack does not
exhibit the characteristic low-entropy pattern assumed by
STRIP, limiting its ability to reliably distinguish
trigger-bearing inputs from clean inputs in either the latent or
activated state.

\end{document}